\documentclass[12pt]{article}
\pdfoutput=1
\usepackage{textcomp}
\usepackage{color}
\usepackage{putex}
\usepackage{autobreak}
\usepackage{indentfirst}
\usepackage{graphicx}
\usepackage{float}
\usepackage{bbold}
\graphicspath{{plots/}}
\usepackage{tabularx}
\usepackage{caption}
\usepackage{amsmath}

\numberwithin{equation}{section}
\usepackage{cancel}
\usepackage{array}
\usepackage{subcaption}
\usepackage{epstopdf}
\usepackage{enumerate}
\usepackage{cite}
\usepackage{youngtab}
\usepackage{tensor}
\usepackage{slashed}
\usepackage[aligntableaux=center]{ytableau}
\usepackage[utf8]{inputenc}
\usepackage{rotating}
\usepackage{adjustbox}
\usepackage{multirow}
\usepackage[
      colorlinks=true,
      linkcolor=red,
      urlcolor=red,
      filecolor=black,
      citecolor=red,
      ]{hyperref}
\usepackage[noabbrev]{cleveref}
\usepackage{tikz, pgfplots}
\usetikzlibrary{arrows.meta, calc, positioning}
\usetikzlibrary{decorations.pathmorphing}

\newcommand{\case}[1]{%
  \sbox0{\bfseries Case #1 }%
  \noindent\hangindent=\wd0\box0\ignorespaces
}

\newcommand{\theorem}[1]{%
  \sbox0{\bfseries Theorem #1 }%
  \noindent\hangindent=\wd0\box0\ignorespaces
}

\newcommand{\step}[1]{%
  \sbox0{\bfseries Step #1 }%
  \noindent\hangindent=\wd0\box0\ignorespaces
}

\newcommand{\cc}[1]{%
  \sbox0{\bfseries ($c_{#1}$) }%
  \noindent\hangindent=\wd0\box0\ignorespaces
}

\newcommand{\cci}[1]{%
  \sbox0{\bfseries ($\infty_{#1}$) }%
  \noindent\hangindent=\wd0\box0\ignorespaces
}
\begin{document}
\institution{duomocathedral}{INFN, Sezione di Firenze; Via G. Sansone 1; I-50019 Sesto Fiorentino (Firenze), Italy}
\institution{forbiddencity}{Kavli Institute for Theoretical Sciences, University of Chinese Academy of Sciences, \cr Beijing 100190, China.}
\title{
Evidence for a $4$-dimensional $\mathcal{N}=1$ integrable quiver in massive type IIA
}
\authors{
Jos\'e Manuel Pen\'in\worksat{\duomocathedral}${}^{,}$\footnote{{\hypersetup{urlcolor=black}\href{mailto:jmanpen@gmail.com}{jmanpen@gmail.com}}} and 
Konstantinos C.~Rigatos\worksat{\forbiddencity}${}^{,}$\footnote{{\hypersetup{urlcolor=black}\href{mailto:92konstantinos10@gmail.com}{rkc@ucas.ac.cn}}}
}
\abstract{
We are examining a newly discovered parametric family of backgrounds in massive type IIA supergravity that contains a warped $\text{AdS}_5$ factor. This family is the dual gravity description of $4$-dimensional quivers with $\mathcal{N}=1$ supersymmetry. We are interested in the status of classical integrability in these theories and we show that there exists a single choice of solutions that is special, while all other choices lead to non-integrable quivers and chaotic string motion. By focusing on this special choice we provide strong suggestive evidence for the integrability of the dual field theory based on analytic studies and extensive numerical analysis. 
}
\date{\today}
\maketitle
{
\hypersetup{linkcolor=black}
\tableofcontents
}
\newpage
\section{Introduction}\label{sec: prologue}
\subsection{Motivation}\label{sec: Motivation}
Integrability is one of the cornerstones in our modern understanding of field theories. Integrable theories have a wealth of conserved quantities, and more specifically Liouville integrability is the statement that the motion of a dynamical system is confined to a submanifold of smaller dimensionality than its phase-space. Not only that, but integrability is, also, a statement that the theory is exactly solvable for any value of the coupling constant. It should be obvious that these theories can be considered as beautiful toy-models to provide us with intuition, since they admit exact and analytic solutions, on the structures of observables in the more interesting and challenging physics theories; see \cite{Arutyunov:2021ygo} for a more detailed exposition. 

Using the holographic dictionary \cite{Maldacena:1997re} we can relate a quantum field theory to the world-sheet description of string theory in an appropriately chosen background. This has a twofold importance in the framework of integrability. On the one hand, we can attempt to phrase the question of integrability in either side depending on which one appears to be the most tractable and on the other, integrable structures in string theory have a role on their own, while also leading to new integrable gauge theories \cite{Beisert:2010jr,Torrielli:2016ufi,Zarembo:2017muf}. 

Even in the most basic and well-founded example of the AdS/CFT duality, the one that relates type IIB superstring theory in $\text{AdS}_5 \times \text{S}^5$ to the four-dimensional $\mathcal{N}=4$ super Yang-Mills theory, the most prominent calculations rely on the integrability of the setup, see \cite{Beisert:2010jr} for a detailed account. Integrability of the classical string description, in this example, is manifest since we can obtain the equations of motion that result from the Lagrangian description as a condition of flatness on the so-called Lax connection \cite{Bena:2003wd}. 

In spite of integrable theories having many coveted features, it is extremely cumbersome to declare a theory as being integrable, even at the classical level. The reason is that the proof of integrability relies on the existence of a Lax connection defined on the cotangent bundle of the theory. The sad reality is that there is no algorithmic and systematic way to construct that quantity. To make matters even worse, the precise statement is that there is no reason to determine, a priori, whether such a connection exists or not. Owing to this discussion, in order to generate integrable theories we resort, by and large, to appropriate deformations which preserve some structures of theories that are known to be integrable \cite{Sfetsos:2013wia,Borsato:2016pas,Delduc:2014kha}.  

Even in two-dimensional models, which is some sense can be considered as the simplest ones, we lack a systematic approach to characterize the status of integrability. Indeed, hitherto we do not posses a classification of two-dimensional integrable $\sigma$-models, at least in all generality. 

While this might seem to be discouraging at the very least, there has been a lot of progress in discovering new integrable theories with explicitly known supergravity description that goes beyond the basic $\text{AdS}_5$/$\text{CFT}_4$ example mentioned above. 

The $\text{AdS}_3 \times \text{S}^3 \times \mathcal{M}^4$, with $\mathcal{M}^4 = \text{T}^4$ or $\mathcal{M}^4 = \text{S}^3 \times \text{S}^1$ enjoys integrability\cite{Babichenko:2009dk}. Furthermore, within the Gaiotto-Maldacena backgrounds \cite{Gaiotto:2009gz}, there exists a special choice, namely the Sfetsos-Thompson solution \cite{Sfetsos:2010uq,Lozano:2016kum}, that is classically integrable \cite{Borsato:2017qsx}. Another example that enjoys integrability is coming from the $\text{AdS}_7$/$\text{CFT}_6$ paradigm and the massive type IIA solution is of the form $\text{AdS}_7 \times \text{S}^2 \times \mathcal{I}_z$ \cite{Filippas:2019puw}. And the final explicitly known gauge/gravity dual pair for which we have a definite statement of classical integrability is the marginal Leigh-Strassler $\beta$-deformation, when the deformation parameter is real. Its dual gravity description is given in terms of the Lunin-Maldacena solution \cite{Lunin:2005jy} and the equivalence of the classical equations of motion of strings to the flatness of the Lax connection was derived in \cite{Frolov:2005dj}. We should note, however, that there is a breakdown of integrability when the deformation parameter is complex \cite{Roiban:2003dw,Frolov:2005ty,Berenstein:2004ys,Giataganas:2013dha}. 

From the list above it is obvious that there is only one example of a classically integrable field theory in four dimensions and with minimum amount of supersymmetry, that of the real Leigh-Strassler deformation. This motivated us to look afresh for a new theory exhibiting these properties.  

The story unfolds as follows: recently in \cite{Merrikin:2022yho} the authors derived twisted compactifications of the $\text{AdS}_7$ solutions developed in the series of papers \cite{Apruzzi:2014qva,Apruzzi:2015wna,Passias:2015gya,Apruzzi:2017nck,Macpherson:2016xwk,Bobev:2016phc,Cremonesi:2015bld} down to $\text{AdS}_5$. This is an example of a flow across dimensions that involves a twisted
compactification. More specifically, the starting point was a family of $6$-dimensional $\mathcal{N}=(1,0)$ superconformal field theories parameterised by a function\footnote{In principle there are infinite choices for the function that parameterises the field theory. In an abuse of language, we sometimes refer to quivers of this type as an ``infinite family".} and after their compactification on a two-dimensional manifold with constant curvature, at the end-point of the flow there exists a new one-parameter family of strongly coupled $4$-dimensional $\mathcal{N}=1$ superconformal field theories.  

Regarding the statement of integrability: in the mother $\text{AdS}_7$ we know that all, but a very specifically chosen, solutions are non-integrable. For the special choice we are certain on its classical integrability since there exists an explicit Lax connection. While one might be tempted to argue that the same question is trivial in the daughter theories, this is far from being true. The twisted compactification that was performed in \cite{Merrikin:2022yho} was such that in the resulting theory, in the $\text{AdS}_5$ backgrounds, only $\mathcal{N}=1$ supersymmetry is preserved. Since the twisted compactification of a given background is not a known method that preserves the integrability structures, whether we can discover signatures of integrability, let alone a proof of it, constitutes a highly non-trivial question. 
\subsection{The approach we undertake in this work}\label{sec: ourwork}
In light of the objective difficulties in spotting integrable structures, that we described above, and combining this with the fact that integrability has to be manifested in a universal manner for a given theory, reverse-engineering the logic of the task at hand and hence searching for non-integrable signatures appears as a ratiocination. 

To understand this dialectic approach to these matters, let us consider classical Hamiltonian systems. There, analytic non-integrability is employing the use of Galois theory on differential equations, through which we can derive a conclusive statement on the structure of these systems. The statements of differential Galois theory on second-order, ordinary, linear differential equations were transformed to much simpler and straightforward algebraic form. Not only that, but also there exists an explicit algorithm producing the Liouvillian solutions of such equations, if any exist. This is the so-called Kovacic's algorithm \cite{KOVACIC19863}.

Now, we turn our attention to supergravity backgrounds. In order to perform the full-fledged analysis of the dynamics of classical strings, we would have to consider a generic string embedding and study the system of non-linear partial differential equations that arise from the string $\sigma$-model. A much more manageable approach is to study certain wrapped string embeddings, which are chosen in accord with the isometries of the background, and then analyse the equations of motion that result from that particular embedding. Not only that, but also these string embeddings are chosen in such a way that they result in ordinary differenetial equations. This is a crucial requirement, in order to be able to apply the criteria of Kovacic's algorithm. 

Since integrability is a property that has to be manifested universally in a theory, it means that each and every single string configuration has to result to integrable dynamics. Ergo, a single counter-example that is echoing signatures of non-integrability is sufficient to allow us to declare the theory as being non-integrable without any loss of generality. This is, in a nutshell, the method that has been dubbed \textit{analytic non-integrability}; see \cite{Basu:2011di} for the first application in the critical string theory context. 

We review the nuts and bolts of the above of the aforementioned method. To begin with, we write a string soliton that has $\mathcal{Z}$ degrees of freedom and derive its equations of motion. Then, we proceed by finding simple solutions for $(\mathcal{Z}-1)$ equations of motion.  These simple solutions define the so-called invariant plane of solutions. We focus on the final equation of motion, and we replace the solutions found previously. Subsequently, we consider linearised fluctuations around this solution. In doing so, we arrive at a second-order, linear differential equation, which is the  so-called the Normal Variational Equation (NVE) and assumes the schematic form $\mathcal{A}_1 f^{\prime \prime} + \mathcal{A}_2 f^{\prime} + \mathcal{A}_3 f = 0$; see also \cite{Rigatos:2020hlq}.

In passing, and for completeness, we would like to mention that S-matrix factorization on the worldsheet theory of the string has been proven to be an invaluable tool to provide us with certain conditions of non-integrability \cite{Wulff:2019tzh,Wulff:2017lxh,Wulff:2017vhv,Wulff:2017hzy}. There is a nice connection between the analytic non-integrability and S-matrix factorization approaches \cite{Giataganas:2019xdj}. 

In the spirit of looking for signatures of non-integrability, an equally important role is played by the standard arrows in our quivers when studying the dynamical evolution of classical systems. That is, we can look for characteristic behaviours of a given system that are indicators of chaos. This can be implemented numerically in order to study quantities such as Lyapunov exponents, Poincar\'e sections, and power spectra. 

At this point it is worthwhile pausing for a moment to mention a subtle point. It has been noted in the past, see for example \cite{Nunez:2018ags,Rigatos:2020hlq}, but we stress it for clarity. While it is not necessary that non-integrability implies chaos, the converse is always true. More specifically, any chaotic system is non-integrable by default. We would, also, like to point out that to the extent of our knowledge, in the context of the gauge/gravity duality, all systems that have been studied and found to be non-integrable are chaotic as well. We are not certain what this is telling us about the dynamics of strings.  

The ideas described above and dubbed analytic non-integrability and studies of chaotic motion were first utilized, in a critical string theory setup, by Basu and Pando-Zayas in \cite{Basu:2011di}, building upon earlier studies of string solitons in the AdS-Schwarzschild \cite{PandoZayas:2010xpn}. Since then there has been many studies of various backgrounds using these ideas. We provide the readers a non-exhaustive, but representative list of related works \cite{Basu:2011fw,Nunez:2018qcj,Nunez:2018ags,Rigatos:2020hlq,Rigatos:2020igd,Filippas:2019ihy,Filippas:2019bht,Stepanchuk:2012xi,Chervonyi:2013eja,Ishii:2021asw,Pal:2021fzb,Roychowdhury:2021jqt,Alencar:2021ljc,Xie:2022yef,Shukla:2023pbp,Kozaki:2023vlu,Xie:2023tjc,Pal:2023kwc,Emelin:2023mwy,Pal:2023bjz,Dutta:2023yhx,Pal:2023ckk}\footnote{It is worthwhile to, also, mention the very interesting approach of \cite{McLoughlin:2022jyt} that examined spectrum of anomalous dimensions and spotted signatures of chaos.}.
\subsection{What this method does and what it does not do}\label{sec: doordonot}
As we have said, the only way to prove integrability is to write the Lax connection \cite{Arutyunov:2021ygo}. An alternative way would be to connect directly to structures which are known to be integrable. However, since the latter structures are known to be integrable, effectively one obtains the Lax connection like so. 

The methods that we use in this work, both the analytic and the numeric, can never prove integrability in a string background. Not even in a single sub-sector of a given theory. As formal proof goes these methods are well-developed and very potent to prove non-integrability; see also \cite{Nunez:2018ags,Filippas:2019bht,Rigatos:2020hlq,Rigatos:2020igd} for related explanations of the method. 

In the case of absence of non-integrability and/or chaos the proper statement to be made is along the following lines: 

\textbf{After having carefully scrutinized all dynamical aspects of the system under examination, we find very strong suggestive evidence that the theory appears to be classically integrable.}

The simple counter-arguments against more powerful and robust statements are the following: 
\begin{itemize}
    \item In case of not spotting non-integrable behaviour analytically: 
    \begin{enumerate}
        \item The string soliton does not capture the breaking of the isometries, and thus it appears we are describing a more symmetric situation than it actually is. 
        \item The string soliton is only chosen such that it leads to ordinary differential equations in order to apply Kovacic's algorithm. However, this is not the most general string embedding.  
    \end{enumerate}
    \item In the absence of chaos: 
        \begin{enumerate}
        \item We have already mentioned that non-integrability does not necessarily imply chaos. Going to the other direction, this means that while a system is not chaotic it could be non-integrable. Hence, the absence of chaotic signatures is not enough to declare a theory integrable.
        \end{enumerate}
\end{itemize}
\subsection{Comments on the field theory realisation}\label{sec: field_theory}
We have already explained the main steps of which our method consists. It is clear that we need to consider a string soliton and study the resulting dynamics. Doing so, we end up obtaining an equation of motion for the time coordinate of the background spacetime. This has to be of appropriate form, namely it has to give the energy of the string as its first integral. While this step might be characterised by most as nitpicking it is highly importance, in order to be able to argue, using the principles of the AdS/CFT duality, that a specific string configuration is a dual gravity description of a gauge invariant operator in the field theory \cite{Filippas:2019bht}. This is necessary in order to be able to argue that the statement of (non)-integrability is shared between the bulk and boundary pictures. With that in mind, we will always incorporate the time coodinate on the worldsheet in the target space time coordinate by means of $t=t(\tau)$ and this is what enters the dynamics through the equations of motion derived from the $\sigma$-model. Having said that our string configuration possesses a well-defined holographic realization; see also \cite{Rigatos:2020igd,Filippas:2019bht} for related comments in different backgrounds.

To put it in simple terms, when resorting to the methods of analytic and numerical (non)-integrability we ought to be extremely careful about the consistency of our string embedding for any given (class of) supergravity background(s). It should be obvious from this discussion that the exact knowledge of the form of the operator is not striclty necessary. The same statement is true for the boundary superconformal field theory as well; see \cite{Rigatos:2020igd} for an application of the method without the knowledge of the specifics of the dual field theory. There is absolutely no need for the precise holographic dictionary to share the statement of non-integrability so long as our string soliton is consistent in the way we described\footnote{This is up to the implicit assumption that the supergravity description has a well-defined global form.}. 

Having already taken a small detour to discuss some aspects of the field theory interpretation, we also feel the need to add some commentary on the qualitative form of the boundary operators. We are using extended string configurations that wrapped in a non-trivial manner along some of the cyclic coordinates of the background. Using the standard holographic dictionary, the field theory operators that are the dual descriptions to these bulk states are long, unprotected operators with large quantum numbers; for example they have large energy and/or angular momentum. In the case of the Klebanov-Witten model this has been more thoroughly elaborated in \cite{Basu:2011di}, since we have a thorough understanding and control over the boundary superconformal field theory. In the limiting case where we take the wrapping of the string to be trivial, we have the low-energy description of the string which appears as a point-particle moving along some geodesic paths.
\subsection{Gaining intuition on discovering integrable theories}\label{sec: intuition}
In the preceding discussion we have clarified what the role and main steps of the analytic and numerical analyses of non-integrability are in the context of examining the classical integrability of string motion. We have not pointed out, however, some intuition that has been built so far based on these studies that allow one to investigate more promising situations. In our opinion, this is the most important and exciting aspect in this kind of studies. 

As we have mentioned, in this work - as in many of the previous ones in the literature - we study a family of supergravity solutions that has non-trivial warp factors. Each of these warp factors can be written in terms of a function, which we call $\alpha(z)$ to connect with the rest of the work, of a single-variable. That variable is one of the internal dimensions of the background metric. 

It was observed in the past, that in situations where the warp factor in front of the $\text{AdS}$ part of the geometry can be set to be equal to a constant consistently, we are led to obtain a classically integrable solution. This is handwavy and intuitive way to look for new integrable holographic theories, and is by no means a golden hammer\footnote{A simple counter-example is the $\text{AdS}_5 \times \text{SE}_5$ solution, with $\text{SE}_5$ being any Sasaki-Einstein space, that is non-integable \cite{Basu:2011di,Basu:2011fw,Rigatos:2020hlq} in spite of having a trivial warp factor in front of $\text{AdS}$.}. Examples that follow this rule are \cite{Borsato:2017qsx,Filippas:2019puw}. 

So, why is the above working and on what premise? The answer, in simple terms, is that by undoing the warping in front of $\text{AdS}$, one is hoping to end up with a background whose geometry is the direct product of integrable sub-spaces. 
\subsection{The structure of this paper}\label{sec: structure}
The structure of this work is the following: in \cref{sec: massive_iia} we review and describe the basic aspects of the supergravity solutions that are of interest to us. We proceed to derive the equations of motion that govern the dynamics in this setup in \cref{sec: closed_strings}. In \cref{sec: special_iia} we present the supergravity solution that appears to be special in the context of classical integrability. \Cref{sec: numerics} contains the results of our numerical approaches. We conclude in \cref{sec: epilogue}.

We have included \cref{app: kovacic} in an effort to make this paper clear and self-contained. It contains the steps of Kovacic's algorithm. In the physics context, the basic steps of Kovacic's algorithm were presented most prominently in \cite[appendix A]{Filippas:2019ihy}. We, however, opt to present Case III of Kovacic's algorithm, which has not been presented before in the physics literature\footnote{To the extent of our knowledge.}. We feel that it is important to mention the excellent reference \cite{Nunez:2018qcj} that offers exposition to both the group theoretic and also  differential equation approach as well.  
\section{Preliminaries}\label{sec: massive_iia}
We provide a brief overview of the main aspects of the new infinite family of massive type IIA backgrounds that is the focus in this work. This solution was obtained in \cite{Merrikin:2022yho} and is a flow between a family of $6$-dimensional $\mathcal{N}=(1,0)$ and $4$-dimensional $\mathcal{N}=1$ superconformal field theories. Based on the existence of a monotonic quantity that interpolates between the conformal points at high- and low-energies there is a proposal for an appropriate quiver capturing the low-energy dynamics. These quivers were proposed to hit a conformal fixed-point at low-energies. In essence it is their strongly-coupled dynamics that is being described by the new infinite family of massive type IIA vacua with an $\text{AdS}_5$ factor. 

It is worthwhile to comment on the following subtlety that was observed in \cite{Merrikin:2022yho}: the $\beta$-functions and the R-symmetry anomalies of the proposed quiver are cancelled. In addition to that, the free-energy scaling with the parameters of the quiver matches the result obtained from the holographic calculation. However, the precise free-energy coefficient in the field theory calculation does not match, exactly, the supergravity prediction.

The physical interpretation of the above comment is that the proposed quiver theory is a good first step to the end of having the appropriate field theory dual of the infinite family of massive type IIA vacua, but improvements need to be made for a precise identification. This, however, does not affect the integrability analysis from the supergravity side that we undertake in this work.

\subsection{The massive type IIA backgrounds}\label{sec: massive_iia_des}
We proceed to present the solution that was obtained in \cite{Merrikin:2022yho} and will be the focus of our analyses. We stress that we are not studying the flow of the $6$-dimensional quivers, but rather we focus on the IR fixed point of the flow, which was obtained in \cite{Passias:2015gya}. As we have already discussed we are interested in the status of integrability in the $4$-dimensional theory.

The NS-NS sector of the solutions consists of the string-frame metric, the dilaton, and a non-trivial $2$-form field. The metric is 
\begin{equation}\label{eq: geometry}
ds^2 = f_1 ds^2_{\text{AdS}_5} + f_2 ds^2_{\mathbb{H}^2} + f_3 dz^2 + f_4 \left( ds^2_{S^2} + \cosh\theta_1 \sin^2 \theta_2 ds^2_2 \right)		\,			,
\end{equation}
with the various submanifolds that appear in \cref{eq: geometry} being given by: 
\begin{equation}
	\begin{alignedat}{2}
&ds^2_{\text{AdS}_5}						=			-\cosh^2 \rho dt^2 + d\rho^2 + \sinh^2 \rho ds^2_{S^3}											\,			,			\quad
&&ds^2_{\mathbb{H}^2}						=			d\theta^2_1 + \sinh^2\theta_1 d\phi^2_1															\,			,			\\
&ds^2_{S^3}								=			d \omega^2_1 + \sin^2 \omega_1 \left( d \omega^2_2 + \sin^2 \omega_2 d \omega^2_3\right)			\,			,			\quad
&&ds^2_{S^2}								=			d\theta^2_2 + \sin^2\theta_2 d\phi^2_2																\,			,			\\
&ds^2_2										=			\cosh\theta_1 d\phi^2_1 + 2 d\phi_1 d\phi_2														\,			.			
	\end{alignedat}
\end{equation}

The dilaton and the NS-NS 2-form are given by: 
\begin{equation}\label{eq: dilaton_b2_field}
e^{4 \Psi} = f_5		\,		,		\quad		B_2 = \left(f_6 \sin\theta_2 d\theta_2 - \pi \cos \theta_2 dz \right) \wedge \left(d\phi_2 + \cosh\theta_1 d\phi_1 \right) 	
\end{equation}

The RR-sector of the theory is not relevant for our studies, however, we provide for completeness the relevant expressions below:
\begin{equation}\label{eq: rr}
	\begin{alignedat}{2}
&F_0		&&=			f_7																																		\,				,	\\
&F_2		&&=			f_6 f_7 \vol_{S^2_c} +	f_8	\left( \cos\theta_2 \vol_{\mathbb{H}^2} - \vol_{S^2_c}  \right)											\,				,	\\
&F_4		&&=			f_9	\cos\theta_2 \vol_{\mathbb{H}^2} \wedge \vol_{S^2} + f_{10} \sin^2 \theta_2 dz \wedge d\phi_2 \wedge \vol_{\mathbb{H}^2}		\,				,	
	\end{alignedat}
\end{equation}
and in all of the above expressions the functions, $f_i$ for the various values of $i$ are functions of $z$ and their explicit form is: 

\begin{equation} \label{eq: def_fs}
	\begin{alignedat}{3}
	& 	f_1 = 12 \pi \sqrt{\frac{6}{m}} \sqrt{- \frac{\alpha}{\alpha^{\prime \prime}}} 																				\,		,	\quad 
	&& 	f_2 = \pi \sqrt{\frac{6}{m}} \sqrt{- \frac{\alpha}{\alpha^{\prime \prime}}}            																		\,		,	\qquad 
	&& 	f_3 = \pi \sqrt{\frac{3m}{2}} \sqrt{- \frac{\alpha^{\prime \prime}}{\alpha}}																					\,		,	\\
	& 	f_4 = \pi \sqrt{6m} \frac{\sqrt{-\alpha^{\prime \prime} \alpha^{3}}}{3 m \alpha \alpha^{\prime \prime}-2{\alpha^{\prime}}^2}			\,		,	\quad 
	&& 	f_5 = \frac{3me^{4 \Psi_0}}{(2 {\alpha^{\prime}}^2-3 m \alpha \alpha^{\prime \prime})} \left(-\frac{\alpha}{\alpha^{\prime \prime}} \right)^3          	\,		,	\\ 
	& 	f_6 = 2 \pi \frac{\alpha \alpha^{\prime}}{2 {\alpha^{\prime}}^2-3 m \alpha \alpha^{\prime \prime}}															\,		,	\quad
	&& 	f_7 = \frac{2^{1/4} e^{-\Psi_0}}{\sqrt{\pi}}\alpha^{\prime \prime \prime}																					\,		,	\\ 
	& 	f_8 = 2^{1/4} \sqrt{\pi} e^{-\Psi_0} \alpha^{\prime \prime}																						          	\,		,	\qquad 
	&& 	f_9 = 2^{5/4} \pi^{3/2} e^{-\Psi_0} \frac{\alpha \alpha^{\prime} \alpha^{\prime \prime}}{2 {\alpha^{\prime}}^2-3 m \alpha \alpha^{\prime \prime}}		\,		,	
 \\
	& 	f_{10} = 2^{1/4}\pi^{3/2}e^{-\Psi_0}\alpha^{\prime \prime}			\,		.
	\end{alignedat}
\end{equation}

In the above and throughout this work we used the abbreviation $\alpha\equiv\alpha(z)$.

The canonical volume forms are given by:
\begin{equation}\label{eq: vols}
    \begin{alignedat}{3}
&\vol_{S^2} 					= \sin \theta_2 d\theta_2 \wedge d\phi_2													\,				,		\quad
&&\vol_{S^2_c} 					= \sin \theta_2 d\theta_2 \wedge\left( d\phi_2 + \cosh \theta_1 d\phi_1	\right)			\,				,		\\
&\vol_{\mathbb{H}^2}			= \sinh \theta_1 d\theta_1 \wedge d\phi_1													\,				.
    \end{alignedat}
\end{equation}

Consistency with the BPS equations and the Bianchi identities that govern the fluxes requires that the mass-parameter, $F_0$, should be piecewise constant. Interpreting the solutions presented above in terms of fully localised sources, D$8$-branes, dictates that $\alpha^{\prime \prime \prime}$ must be piecewise constant. It is precisely the various consistent possibilities from which we can choose the function F$_0$ that provides us with a parametric family of solutions. It is should clear from the previous comments that $\alpha''' \propto F_0$. 

We note that in above, we have used the abbreviation ${}^{\prime} \equiv \partial_z$ and we keep this shorthand notation throughout our work. 

\subsection{The quiver picture}
The authors of \cite{Merrikin:2022yho} put forth a phenomenological proposal in terms of which we can think the dual superconformal field theories. In order to have a better understanding of the systems under consideration in this work, we review some of their main characteristics.  

We begin by pointing out that in the UV regime the system above is dual to a $6$-dimensional superconformal field theory. More precisely, one can explicitly check that the solutions asymptote to a background spacetime that is of the form $\text{AdS}_7 \times \mathcal{I}_z \times \text{S}^2$. Then, the $6$-dimensional field theory is understood to be defined on a spacetime of the form $\mathbb{R}^{1,3} \times \mathbb{H}_2$. This and the crucial fact that there exists a non-trivial fibration of the $\mathbb{H}^{2}$ over the $S^{2}$ of the internal manifold that does not vanish at infinity amounts to precisely provide the topological twisting in the field theory such that it partially preserves some amount of supersymmetry.

The above line of reasoning and specifically the parallelism to a $6$-dimensional superconformal field theory leads one to suggest naturally a geometric engineering in terms of magnetically charged $5$-branes, NS$5$-branes, with D$6$- and D$8$-branes. Of course, this argument is rough around the edges. 

The dual field theory has colour (gauge) and flavour nodes. The associated numbers characterising the nodes are $N_1, N_2, \ldots , N_P$ and $F_1, F_2, \ldots , F_P$ respectively. And now we have to turn to the $6$-dimensional picture in order to make a connection amongst those numbers explicitly. More specifically, anomaly cancellation in the $6$-dimensional quivers requires that \cite{Cremonesi:2015bld}:
\begin{equation}
    F_i = 2 N_i - N_{i-1} - N_{i+1}
\,            ,
\end{equation} 
a condition that is ensured when $\alpha(z)$ is chosen such that its second derivative is piecewise linear and continuous. 

At this point, and since we know the special solution that we will present later and its physical interpetation, we wish to stress that Cremonesi and Tomasiello in \cite{Cremonesi:2015bld} observed that there exists a possible scaling that involves taking the number of D$8$-branes to infinity, and hence creating a continuous (smeared) distribution. However, under this scaling the $\text{AdS}_7$ backgrounds are still trustworthy descriptions of $6$-dimensional quivers with $\mathcal{N}=(1,0)$ and there are no issues with anomalies. As we will see below, this will play a crucial role to our suggestion for the integrable quiver.

Going back to the basic characteristics of the quiver in our setups, we have reached a UV conformal fixed-point. These UV fixed-points are deformed by some vacuum expectation values (VEVs) or by some relevant operators. These
deformations, in close analogy to the presence of fibrations, topologically twist the $6$-dimensional superconformal field theory and trigger a renormalisation group flow, that ends in a $4$-dimensional superconformal field theory.

The Page charge quantisation that was performed in \cite{Merrikin:2022yho} suggests the existence of a new, second set of NS$5$-branes. This is due to the twisting. So, we have exactly $(g-1)P$ NS$5$-branes, on top of the previous one originating in the $6$-dimensional picture, that are orthogonal to the original NS$5$-branes, with $g$ being a number. This generates additional colour groups. So, with this extra stack of NS$5$-branes, the D$6$-branes can stretch between the different stacks of the NS-branes. 

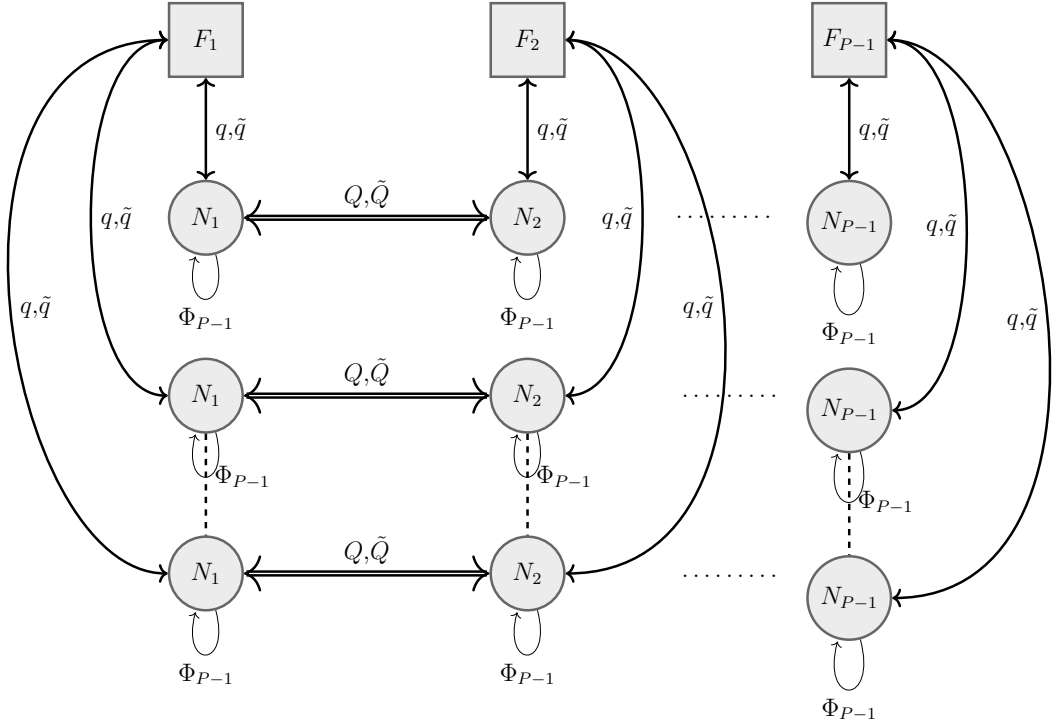
\begin{figure}[H]
\begin{adjustbox}{width=1\textwidth}
\begin{tikzpicture}
[
node distance = 17mm and 41mm,
b/.style={rectangle, draw=black!60, fill=gray!15, very thick, minimum size=35},
c/.style={circle, draw=black!60, fill=gray!15, very thick, minimum size=35}
]

\node[b]      (posFk)                               {$F_{P-1}$};
\node[c]      (posak)      [below=of posFk]         {$N_{P-1}$};
\node[c]      (posmuk)     [below=of posak]         {$N_{P-1}$};
\node[c]      (posFtk)     [below=of posmuk]        {$N_{P-1}$};

\node[b]      (posFF)      [left=of posFk]          {$F_2$};
\node[c]      (posakk)     [below=of posFF]         {$N_2$};
\node[c]      (posmukk)    [below=of posakk]        {$N_2$};
\node[c]      (posFtkk)    [below=of posmukk]       {$N_2$};

\node[b]      (posFFF)     [left=of posFF]          {$F_1$};
\node[c]      (posakkk)    [below=of posFFF]        {$N_1$};
\node[c]      (posmukkk)   [below=of posakkk]       {$N_1$};
\node[c]      (posFtkkk)   [below=of posmukkk]      {$N_1$};

\draw[<->, very thick] (posFk.south) to node[right] {$q$,$\tilde{q}$} (posak.north);
\draw[<->, very thick] (posmuk.east)  .. controls  +(right:17mm) and +(right:17mm) .. (posFk.east) node [left,pos=0.5] {$q$,$\tilde{q}$};
\draw[<->, very thick] (posFtk.east)  .. controls  +(right:43mm) and +(right:27mm) .. (posFk.east) node [left,pos=0.5] {$q$,$\tilde{q}$};

\path (posak)   edge [loop below]  node [below] {$\Phi_{P-1}$} (posak);
\path (posmuk)  edge [loop below]  node [right] {$\Phi_{P-1}$} (posmuk);
\path (posFtk)  edge [loop below]  node [below] {$\Phi_{P-1}$} (posFtk);

\draw[<->, very thick] (posFF.south) to node[right] {$q$,$\tilde{q}$} (posakk.north);
\draw[<->, very thick] (posmukk.east)  .. controls  +(right:17mm) and +(right:17mm) .. (posFF.east) node [left,pos=0.5] {$q$,$\tilde{q}$};
\draw[<->, very thick] (posFtkk.east)  .. controls  +(right:43mm) and +(right:27mm) .. (posFF.east) node [left,pos=0.5] {$q$,$\tilde{q}$};

\path (posakk)   edge [loop below]  node [below] {$\Phi_{P-1}$} (posakk);
\path (posmukk)  edge [loop below]  node [right] {$\Phi_{P-1}$} (posmukk);
\path (posFtkk)  edge [loop below]  node [below] {$\Phi_{P-1}$} (posFtkk);

\draw[<->, very thick] (posFFF.south)  to node[right] {$q$,$\tilde{q}$} (posakkk.north);
\draw[<->, very thick] (posFFF.west)     .. controls  +(left:17mm) and +(left:17mm)          .. (posmukkk.west) node [right,pos=0.5] {$q$,$\tilde{q}$};
\draw[<->, very thick] (posFFF.west)     .. controls  +(left:43mm) and +(left:27mm)          .. (posFtkkk.west) node [right,pos=0.5] {$q$,$\tilde{q}$};

\path (posakkk)   edge [loop below]  node [below] {$\Phi_{P-1}$} (posakkk);
\path (posmukkk)  edge [loop below]  node [right] {$\Phi_{P-1}$} (posmukkk);
\path (posFtkkk)  edge [loop below]  node [below] {$\Phi_{P-1}$} (posFtkkk);

\draw[<->,double , very thick] (posakkk.east)     to node[above] {$Q$,$\tilde{Q}$} (posakk.west);
\draw[<->,double , very thick] (posmukkk.east)     to node[above] {$Q$,$\tilde{Q}$} (posmukk.west);
\draw[<->,double , very thick] (posFtkkk.east)     to node[above] {$Q$,$\tilde{Q}$} (posFtkk.west);

\node at (-21mm,-29.5mm) (aux4) {$\bf\textcolor{black}{\ldots \ldots \ldots}$};
\node at (-20mm,-59.5mm) (aux5) {$\bf\textcolor{black}{\ldots \ldots \ldots}$};
\node at (-20mm,-89.5mm) (aux5) {$\bf\textcolor{black}{\ldots \ldots \ldots}$};

\draw[-, very thick, dashed] (posmukkk.south) to node[right] {} (posFtkkk.north);
\draw[-, very thick, dashed] (posmukk.south) to node[right] {} (posFtkk.north);
\draw[-, very thick, dashed] (posmuk.south) to node[right] {} (posFtk.north);

\end{tikzpicture}
\end{adjustbox}
\caption{
The quiver field theory. The $\Phi_i$  adjoint fields obtain a mass. We have used a double arrow and $Q$,$\tilde{Q}$ for the bi-fundamentals fields, while for the fundamentals are denoted by $q, \tilde{q}$. The vector multiplets are represented by the circles. We have used a $4$-dimensional, $\mathcal{N}=1$ language. In this diagram, the $\ldots \ldots \ldots$ should be understood as continuing the quiver in the same pattern; e.g bi-fundamentals connect $N_2$ to $N_3$, and $N_3$ to $N_4$ and so on. 
}
\label{fig: quiver}
\end{figure}

After a careful counting of the degrees of freedom in the presence of the new ingredients, it was suggested that a quiver that consists of $P^2 (g-1)$ copies of the $6$-dimensional mother theory should correctly capture the essential aspects in the $4$-dimensional daughter theory \cite{Merrikin:2022yho}. The quiver that was suggested in \cite{Merrikin:2022yho} is depicted in \cref{fig: quiver}

In the cartoon of the quiver, \cref{fig: quiver}, we have used the very common notation for quiver diagrams. Namely, arrowed lines denote chiral multiplets, circles stand for vector multiplets, and boxes are flavours.
\section{Dynamics of classical closed strings}\label{sec: closed_strings}
In this section we are going to study analytically the motion of a closed string soliton in the backgrounds presented in \cref{sec: massive_iia}. Such dynamics are governed by the equations of motion derived from the extremization of the string-$\sigma$ model action 
\begin{equation}\label{eq: sigma_action}
S_{\sigma} = \frac{1}{4 \pi \alpha^{\prime}} \int d^2 \sigma \left(h^{\alpha \beta} G_{MN} + \epsilon^{\alpha \beta} B_{MN} \right) \partial_{\alpha} X^M \partial_{\beta} X^N 
\end{equation}
which has to be supplemented by the Virasoro constraint $T_{\alpha \beta}=0$, with:
\begin{equation} \label{eq: energy_momentum_tensor_def}
T_{\alpha \beta} = \frac{1}{\alpha^{\prime}} \left( G_{MN} ~ \partial_{\alpha} X^M \partial_{\beta} X^N - \frac{1}{2} G_{MN} ~ \eta_{\alpha \beta} ~ \eta^{\gamma \delta} ~ \partial_{\gamma} X^M \partial_{\delta} X^N \right)\, ,
\end{equation}
where $G_{MN}$ and $B_{MN}$ are the target space metric and NS field, $X^M$ are the target spacetime coordinates and $d^2\sigma=d\tau d\sigma$, where $(\tau,\sigma)$ are the worldvolume coordinates. In the equations above we choose the convention: $\eta_{\sigma\sigma}=-\eta_{\tau\tau}=1, \eta_{\tau\sigma}=0$, and $\epsilon^{\alpha\beta}$ is the Levi-Civita tensor with convention $\epsilon^{\tau\sigma}=1$. \Cref{eq: sigma_action} makes explicit the fact that the string soliton couples to the target spacetime through a pullback to the worldvolume coordinates. $T_{\alpha\beta}$ is the worldsheet energy-momentum tensor.

The starting point to study integrability analytically is to provide and ansatz for a suitable embedding $X^M(\tau, \sigma)$ capturing the coupling to the NS field and respecting the isometries of the background, and construct the Normal Variational Equation (NVE). Accordingly, we place our string at three fixed angles on the three-sphere inside the $\text{AdS}_5$ and further suggest the following string embedding:
\begin{equation}\label{eq: string_emb}
t=t(\tau)							\,	,		
\rho=\rho(\tau)						\,	,		
z=z(\tau)							\,	,			
\theta_1 = \theta_1(\tau) 			\,	,			
\theta_2 = \theta_2(\tau)			\,	,	
\phi_1 = \alpha_1 \sigma			\,	,			
\phi_2 = \alpha_2 \sigma			\,	.
\end{equation}
For the above embedding, \cref{eq: string_emb}, we evaluate the $\sigma$-model Lagrangian to get
\begin{equation}\label{eq: lag_density}
	\begin{aligned}
\mathcal{L} = 	&f_1\left( \cosh^2 \rho \dot{t}^2 - \dot{\rho}^2 \right)  - f_3 \dot{z}^2  - f_2 \dot{\theta_1}^2 - f_4 \dot{\theta_2}^2 	+ \\
				& \left(f_2\sinh^2 \theta_1 + f_4\cosh^2 \theta_1 \sin^2 \theta_2 \right)\alpha^2_1 + f_4 \sin^2 \theta_2 \alpha^2_2 + 2 f_4 \cosh\theta_1 \sin^2 \theta_2 \alpha_1 \alpha_2 + \\
				&2f_6\sin\theta_2\left(\cosh \theta_1 \alpha_1 + \alpha_2		\right)	\dot{\theta_2} - 2 \pi \cos \theta_2 \left( \cosh \theta_1 \alpha_1 + \alpha_2 \right)	\dot{z}				\,			.
	\end{aligned}
\end{equation} 

The equations of motion that follow from the Lagrangian density given by \cref{eq: lag_density} are: 
\begin{equation}\label{eq: el_eqns}
	\begin{aligned}
\dot{t}					&=		\frac{E}{f_1~\cosh^2 \rho}																																				\,			,		\\
\ddot{\rho}				&=		-\frac{E^2}{f^2_1~\cosh^2 \rho}\tanh\rho - \frac{{f_1}^{\prime}}{f_1} \dot{z} \dot{\rho}																				\,			,		\\
2f_3\ddot{z}			&=		f^{\prime}_1 \left(-\frac{E^2}{f^2_1~\cosh^2 \rho} + \dot{\rho}^2 \right) - f^{\prime}_3\dot{z}^2 - 2 \pi \alpha_1 \sinh \theta_1 \cos \theta_2 \dot{\theta}_1 + 	f^{\prime}_2 \left(\dot{\theta}_1^2 - \alpha^2_1 \sinh^2 \theta_1 \right) + 
\,					\\
&2 \left(\alpha_1 \cosh \theta_1 + \alpha_2 \right)\sin \theta_2 \left(\pi-f^{\prime}_6\right)\dot{\theta}_2 + f_4 \left(\dot{\theta}^2_2 - \left(\alpha_1 \cosh\theta_1 + \alpha_2 \right)^2\sin^2 \theta_2 \right)
\,			,		\\
2f_2\ddot{\theta}_1		&= - 2 f_2 \alpha^2_1 \cosh \theta_1 \sinh \theta_1 -2 \alpha_1 \left(\alpha_1 \cosh \theta_1 + \alpha_2 \right)f_4 \sinh\theta_1 \sin^2\theta_2 + 2 \pi \alpha_1 \sinh \theta_1 \cos \theta_2  \dot{z} 
\,					\\
&-2 f^{\prime}_2 \dot{z} \dot{\theta}_1 -2  \alpha_1 f_6 \sinh \theta_1 \sin \theta_2 \dot{\theta}_2	
\,			,		\\
2f_4\ddot{\theta}_2		&=- \left(\alpha_1 \cosh \theta_1 + \alpha_2\right)^2 f_4 \sin(2 \theta_2) + 2 \alpha_1 f_6 \sinh \theta_1 \sin \theta_2 \dot{\theta}_1 -
\,					\\
&2 \dot{z} \left[ \left( \alpha_1 \cosh \theta_1 + \alpha_2 \right) \sin \theta_2 \left(\pi - f^{\prime}_6 \right) + f^{\prime}_4 \dot{\theta}_2	 \right]	
\,			,
	\end{aligned}
\end{equation}
where $E$ is a constant obtained by integrating the equation of motion for $t(\tau)$ and is associated with the energy of the string.

The equations of motion resulting from the Lagrangian, given by \cref{eq: el_eqns}, are constrained by the Virasoro conditions, namely the worldsheet equations of motion. We can use \cref{eq: energy_momentum_tensor_def} to evaluate the different components of the stress-energy tensor, and thus the Virasoro conditions explicitly for the string embedding given by \cref{eq: string_emb} read: 
\begin{equation}\label{eq: stresstensor}
	\begin{aligned}
T_{\tau \sigma} 	&= T_{\sigma \tau} 			= 0 												\,	,		\\
2T_{\tau \tau} 		&= 2 T_{\sigma \sigma}		= f_1\left(-\cosh^2 \rho \dot{t}^2 + \dot{\rho}^2  \right) + f_2 \dot{\theta_1}^2 + f_4 \dot{\theta_2}^2 + f_3 \dot{z}^2 + \\
					&\left(f_2 \sinh^2 \theta_1 + f_4 \cosh^2 \theta_1 \sin^2 \theta_2 \right)\alpha^2_1 + f_4 \sin^2 \theta_2 \alpha^2_2 + 2 f_4 \cosh \theta_1 \sin^2 \theta_2 \alpha_1	\alpha_2		\,			.
	\end{aligned}
\end{equation}

The constraints above have to hold regardless of the Lagrangian equations of motion. It is easy to show that the energy-momentum tensor, \cref{eq: stresstensor}, is conserved on-shell; evaluated on the equations of motion given by \cref{eq: el_eqns}. More explicitly, we have $\nabla^{\alpha}T_{\alpha \beta}=0$ when we enforce the equations of motion on these expressions. 

It is this agreement, precisely, between the Virasoro conditions and the equations of motion for the target spacetime coordinates, that indicates the consistency of the string soliton chosen in \cref{eq: string_emb}. This serves as a nice and quick check of the string embedding.

We, now, turn to the Hamiltonian formulation of the problem, in order to have a better grip on the physics and subsequent analysis. We begin by presenting the expressions for the conjugate momenta, which are: 

\begin{equation}\label{eq: conjugate_momenta}
	\begin{aligned}
p_t             &=      2 f_1 ~ \cosh^2 \rho ~ \dot{t} 
\,          , 
\\
p_{\rho}        &=      -2 f_1 ~ \dot{\rho}
\,         ,
\\
p_z             &=      -2 f_3 ~ \dot{z} - 2 \pi ~ (\alpha_1 \cosh \theta_1 + \alpha_2) ~ \cos \theta_2
\,          ,
\\
p_{\theta_1}    &=      -2 f_2 ~ \dot{\theta_1}
\,          ,
\\
p_{\theta_2}    &=      - f_4 \dot{\theta_2} + 2 f_6 ~ (\alpha_1 \cosh \theta_1 + \alpha_2) ~ \sin \theta_2
\,          .
	\end{aligned}
\end{equation}
We can, of course, use the expressions in \cref{eq: conjugate_momenta} to derive an explicit expression for the Hamiltonian density. It is given by:
\begin{equation}\label{eq: hamiltonian}
\begin{aligned}
\mathcal{H} = &-f_1 \cosh^2 \rho \dot{t}^2 + f_1 \dot{\rho}^2 + f_3 \dot{z}^2 + f_2 \dot{\theta_1}^2 + f_4 \dot{\theta_2}^2 - 
\\
&\left(f_2\sinh^2 \theta_1 + f_4\cosh^2 \theta_1 \sin^2 \theta_2 \right)\alpha^2_1 - f_4 \sin^2 \theta_2 \alpha^2_2 - 2 f_4 \cosh\theta_1 \sin^2 \theta_2 \alpha_1 \alpha_2
\,              .
\end{aligned}
\end{equation}

In the Hamiltonian formalism of the problem the statement of the Virasoro conditions is equivalent to $\mathcal{H}=0$. The Hamiltonian equations of motion that follow from \cref{eq: hamiltonian} are equivalent to those that follow from the Lagrangian. 

Owing to the above picture, it is very straightforward and instructive to provide a classical mechanics description of the system we have considered. The dynamics of the string embedding has effectively reduced to that of a particle that is moving in the presence of a non-trivial potential. It is precisely the non-trivial wrapping of the string soliton, the non-trivial winding of the string around the cyclic coordinates, that is responsible for generating the potential in the dynamics. The effective mass can be read off from the kinetic terms of the Hamiltonian and is due to the geometrical aspects of the setup we are examining.  

The dynamics of the string soliton in \cref{eq: string_emb} is described by a quite involved system of differential equations, see \cref{eq: el_eqns}. The major reason for the complication is that the equations are inherently coupled. Solving this system in all generality presents itself as a quite formidable task. In spite of this, there is an elegant alternative approach that manages to circumvent this complication and facilitates the needs for the forthcoming analysis. In order to make progress, we need to find a simple solution to the equations of motion and subsequently consider fluctuations around those. Such a fluctuation around the set of simple solutions is what is known as the NVE for a given coordinate. 

Before we proceed, we stress that the Virasoro conditions constitute primary constraints, namely the simple solutions to the equations of motion have to satisfy the Virasoro conditions as well. 

In order to make progress we wish to find an invariant plane of solutions and a simple equation for the $z(\tau)$. To that end, it is easy to check that\footnote{It should, also, be quite obvious that this plane of solutions is by no means unique. It is chosen for our convenience in this analysis, and always taking into consideration the string theory constraints.}
\begin{equation}\label{eq: invariant_plane}
\ddot{\rho} = \dot{\rho} = \rho = \ddot{\theta}_1 = \dot{\theta}_1 = \theta_1 = \ddot{\theta}_2 = \dot{\theta}_2 = \theta_2 = 0 		\,		,
\end{equation}
solves all the equations in \cref{eq: el_eqns} except for the one for $z(\tau)$. We evaluate the $z$-equation on the invariant plane of solution and we obtain 
\begin{equation}\label{eq: simple_z}
2 f_3 \ddot{z} + f^{\prime}_3 \dot{z}^2 + E^2 \frac{f^{\prime}_1}{f^2_1} = 0 	\,	.
\end{equation}
We insert the expressions for the $f_1$ and $f_3$ in the above and we obtain 
\begin{equation}
\ddot{z} - \left( \frac{\alpha \alpha^{\prime \prime \prime} - \alpha^{\prime} \alpha^{\prime \prime}}{4 \alpha^2}\right) \left( \frac{\alpha}{\alpha^{\prime \prime}} \right) \left( \frac{E^2}{36 \pi^2} - \dot{z}^2 \right)  = 0 			\,			,	
\end{equation}
which admits the simple solution 
\begin{equation}\label{eq: z_simple_sol}
z_{\text{sol}}(\tau) = \frac{E}{6 \pi} \tau			\,				.
\end{equation}

It is a very straightforward calculation to evaluate the Virasoro conditions, \cref{eq: stresstensor}, on the invariant plane of solutions given by \cref{eq: invariant_plane} and subsequently enforce \cref{eq: z_simple_sol} to check the agreement of these expressions. 

Therefore, we have shown that all the steps, so far, are self-consistent and allow for a natural interpretation of the string embedding as some operator with large R-charge and large conformal dimension in the dual field theory. 

We wish to make a comment about the invariant plane of solutions around which we fluctuate the string embedding. Naively, it would seem that the choice made in \cref{eq: invariant_plane} reduces the string to a point-particle and hence the effect of the winding of the modes is neglected. However, this is not true as can be seen explicitly from the \cref{eq: NVEtheta1,eq: NVEtheta2}.

This sets up the scene for the derivation of the NVEs for the various dimensions. The NVEs play the protagonistic role in our analysis. 

Since we have found a base solution for our system, we wish to study the NVE. This is achieved by allowing for small fluctuations in one of the equations of motion, working in linear order and evaluating the resulting expression on the base solution defined by the invariant plane of solutions and the simple solution for $t$ and $z$. All the NVEs can be brought into the schematic form 
\begin{equation}
\ddot{y} + \mathcal{A}_1 \dot{y} + \mathcal{A}_2 y = 0			\,			.
\end{equation}
and under the change of variables given by $y = e^{\tfrac{1}{2} \int d \tau \mathcal{A}_1 } q$ we obtain the Schr\"odinger form:
\begin{equation}
\ddot{q} = \mathcal{V} q		\,	,	\qquad	\mathcal{V} = \frac{1}{4}\left(2 \mathcal{A}^{\prime}_1 +\mathcal{A}^2_1  - 4 \mathcal{A}_2 \right)				\,				,
\end{equation}
where in the above $y$ denotes any generic fluctuation.

We will specialise the above picture to each one of the different coordinates and derive their associate NVEs. 
\subsection*{NVE for \texorpdfstring{$\rho$}{rho}}
Here, we derive the NVE for the $\rho$-coordinate, by considering $\rho = 0 + \epsilon r$ and expanding in the $\epsilon \rightarrow 0$ limit working to linear order in the small parameter. This results in:
\begin{equation}
     r''(\tau)+\mathcal{A}_1 r'(\tau)+\mathcal{A}_2r(\tau)=0
\,              ,
\end{equation}
with
\begin{equation}
	\begin{aligned}
&\mathcal{A}_1=\frac{E}{12\pi}\left(\frac{\alpha'}{\alpha}-\frac{\alpha'''}{\alpha''}\right)\bigg\lvert_{z=z_{\text{sol}}}
\,          ,
\\
&\mathcal{A}_2=-\frac{E^2 m}{864\pi^2}\frac{\alpha''}{\alpha}\bigg\lvert_{z=z_{\text{sol}}}
\,          .
	\end{aligned}
\end{equation}
We perform the appropriate change of variables as described above to bring it in the form:
\begin{equation}\label{eq: NVErho}
	\begin{aligned}
\ddot{\varrho} &= \mathcal{V}_{\varrho} ~ \varrho		\,			,		\quad		
\mathcal{V}_{\varrho} &= - \frac{E^2}{f^2_1} \left(1+ \frac{1}{144\pi^2} {f^{\prime}_1}^2 -\frac{1}{72\pi^2 }  f_1 f^{\prime \prime}_1  \right)		\,		.
	\end{aligned}
\end{equation}
	\subsection*{NVE for \texorpdfstring{$\theta_1$}{theta1}}
Likewise, we consider $\theta_1 = 0 + \epsilon v_1$, and work as before. We have:
\begin{equation}
v_1''(\tau)+\mathcal{A}_1v_1'(\tau)+\mathcal{A}_2v_1(\tau)=0
\end{equation}
with
\begin{equation}
	\begin{aligned}
&\mathcal{A}_1=\frac{E}{12\pi}\left(\frac{\alpha'}{\alpha}-\frac{\alpha'''}{\alpha''}\right)\bigg\lvert_{z=z_{\text{sol}}}
\,          ,
\\
&\mathcal{A}_2=\alpha_1^2-\frac{\alpha_1E\sqrt{m}}{6\sqrt{6}\pi}\sqrt{-\frac{\alpha''}{\alpha}}\bigg\lvert_{z=z_{\text{sol}}}
\,          .
	\end{aligned}
\end{equation}
In this case as well, we wish to manipulate the NVE and bring it into a Schr\"odinger form as follows:
\begin{equation}\label{eq: NVEtheta1}
	\begin{aligned}
\ddot{\vartheta_1} &= \mathcal{V}_{\vartheta_1} ~ \vartheta_1		\,			,				\\		
\mathcal{V}_{\vartheta_1} &=-\frac{1}{f_2^2}\bigg(\alpha_1^2f_2^2-\frac{\alpha_1E}{6}f_2+\frac{E^2}{144\pi^2}f_2'^2-\frac{E^2}{72\pi^2}f_2f_2''\bigg)
	\end{aligned}
\end{equation}
	\subsection*{NVE for the \texorpdfstring{$\theta_2$}{theta2}}
Finally, we derive the NVE associated with the $\theta_2$-dimension. To do so, we allow $\theta_2 = 0 + \epsilon v_2$ and proceed as in the previous cases. This yields: 
\begin{equation}
    v_2''(\tau)+\mathcal{A}_1 v_2'(\tau)+\mathcal{A}_2 v_2(\tau)=0
\,              ,
\end{equation}
where in the above we have:
\begin{equation}
	\begin{aligned}
&\mathcal{A}_1=\frac{E}{12\pi(3m\alpha \alpha''-2\alpha'^2)}
\bigg(\frac{\alpha'}{\alpha}((8+3m)\alpha \alpha''-6\alpha'^2)-\frac{\alpha'''}{\alpha''}(2\alpha'^2+3m\alpha \alpha''\bigg)\bigg\lvert_{z=z_\text{sol}} 
\,          ,
\\
& \mathcal{A}_2=(\alpha_1+\alpha_2)^2+\frac{(\alpha_1+\alpha_2)E}{\sqrt{6m}\pi(3m\alpha \alpha''-2\alpha'^2)}
\,  \cdot        
\\
&\bigg(\frac{2(1-3m)\alpha'^2\sqrt{-\alpha^3\alpha''}}{3\alpha^2}-\frac{m\alpha^2((2+3m)\alpha''^2-2\alpha'\alpha'''}{2\sqrt{-\alpha^3\alpha''}}\bigg)\bigg\lvert_{z=z_\text{sol}} 
\,          .
	\end{aligned}
\end{equation}
As we did in the previous cases, we change variables appropriately to bring it to a Schr\"odinger form: 
\begin{equation}\label{eq: NVEtheta2}
	\begin{aligned}
\ddot{\vartheta_2} &= \mathcal{V}_{\vartheta_2} ~ \vartheta_2		\,			,				\\		
\mathcal{V}_{\vartheta_2} &= -  \frac{1}{f^2_4} \left(\vphantom{\frac{1}{2}}  \left(\alpha_1 + \alpha_2 \right)^2  f^2_4 +  \frac{E}{6\pi}\left(\alpha_1 + \alpha_2 \right)  \left(\pi - f^{\prime}_6 \right)f_4 +  \frac{E^2}{144\pi^2} {f^{\prime}_4}^2 - \frac{E^2}{72\pi^2} f_4 f^{\prime \prime}_4 \vphantom{\frac{1}{2}} \right)		\,		.
	\end{aligned}
\end{equation}
We have ended up with second-order, linear, ordinary differential equations. These have all the necessary characteristics to be amenable to Kovacic's algorithm, and hence examine whether or not they admit Liouvillian solutions. 
\section{Liouvillian Integrability}\label{sec: special_iia}
In this section we discuss the status of integrability for closed-strings in the background that we presented in \cref{sec: massive_iia_des}. As we shall see all choices for $\alpha(z)$ consistent with the BPS equations and the Bianchi identities lead to non-integrable theories, except for the solution $\alpha(z) \propto \sin z$, where the proportionality is up to some numerical coefficients. These coefficients from the point of view of $\alpha$ are just the amplitude and the phase of the $sin$ function. From the point of view of the quiver, they control the overal size and the number of flavours. All other solutions for $\alpha(z)$ that are consistent with the BPS equations and Bianchi identities lead to non-integrability. En passant, we make some comments on the intepretation of the special solution.
	\subsection{The allowed solutions for \texorpdfstring{$\alpha(z)$}{alpha}}\label{sec: generic_alpha}
The authors in \cite{Merrikin:2022yho} have thoroughly analysed the BPS equations, Bianchi identities and Page-charge quantisation for the supergravity backgrounds presented \cref{sec: massive_iia}. We review the main results of their analysis in order to understand what are the possible choices for the defining function $\alpha(z)$. We start by recalling that the authors of \cite{Merrikin:2022yho} showed that in order for the solution to be given an interpretation as fully localised D$8$-branes, we must insist on $\alpha^{\prime \prime \prime}(z)$ being piecewise constant, since this is related to the Ramond-Ramond scalar. After performing a gauge transformation, the study of quantisation of the Page charges reveals that $\alpha^{\prime \prime}(z)$ must be a linear function with integer coefficients. Further, they concluded that the fourth-derivative, $\alpha^{(4)}(z)$, should vanish. Finally, in order to avoid singular behaviour the defining function should satisfy $\alpha(0)=0=\alpha(P)$, with $P$ being the upper end of the $z$-interval; $\mathcal{I}_z \in [0,P]$. 

Under these considerations, and following \cite{Merrikin:2022yho}, the most general solution consistent with the supergravity equations is given by:
\begin{equation}\label{eq: alpha_generic}
    \alpha(z) = \alpha_0 + \alpha_1 z + \alpha_2 z^2 + \alpha_3 z^3
    \,              ,
\end{equation}
with $\alpha_{0},\ldots,\alpha_{3}$ being some appropriately chosen constants. 

The authors of \cite{Merrikin:2022yho} also gave a Hanany-Witten interpretation of the setup. The above picture can be realised in terms of a Hanany-Witten setup made of two sets of NS$5$-branes, two sets of D$6$- and one of D$8$-branes. From the $6$-dimensional picture this can be roughly understood as follows. The $\text{AdS}_7$ backgrounds, from which the $\text{AdS}_5$ solutions originate, have an interpretation in terms of a Hanany-Witten setup with NS$5$-D$6$-D$8$ branes \cite{Cremonesi:2015bld} and the twisted compactification induces a new set of NS$5$- and a new set of D$6$-branes. 

Since we have already obtained the NVEs in all generality in the previous section, see \cref{sec: closed_strings}, and we have explained the basic aspects of $\alpha(z)$, we can use the most general and schematic form for the defining function given by \cref{eq: alpha_generic} so that we check if we can make some robust statement on the integrability of closed strings in the background.

	\subsection{Some examples of linear quivers}\label{sec: linearquivers_examples}
To provide the interested reader a firm grip on the shapes of quivers and how they relate to the choice of $\alpha(z)$, we briefly review two examples and make some comments. These were presented thoroughly in \cite{Merrikin:2022yho}. 
\subsubsection*{Example 1}
This first example is the dual originating from a $6$-dimensional liner quiver theory of rank $N_j = j$ and a flavour group given by $\text{SU}(PN)$. The defining function $\alpha(z)$ is given by:
\begin{equation}\label{eq: alpha_exampe1}
    \alpha(z) = 
       \begin{cases}
       \frac{N}{6} (1-P^2)z +\frac{N}{6} z^3,&
        \,           
       0\leq z \leq (P-1)
       \,           ,
       \\
       \\
       -\frac{1}{6} (2P^2-3P+1)(P-z) +\frac{N}{6} (P-1) (P-z)^3, & 
       \,           
       (P-1)\leq z \leq P
       \,           .
     \end{cases}
    \end{equation}
We note that $ \alpha(z)$ vanishes at $z=0$ and $z=P$ and is continuous at $z=(P-1)$. Its derivative, $\alpha'(z)$, is continuous at the same point. 

\subsubsection*{Example 2}
The second example, from a $6$-dimensional point of view, describes a linear quiver with $(P-1)$ gauge nodes of rank $N_j = N$ and flavours $F_J= N(\delta_{J,1}+ \delta_{J, P-1})$. The defining function $\alpha(z)$ in this example is given by:
 \begin{equation}\label{eq: alpha_exampe2}
    \alpha(z) = 
       \begin{cases}
       \frac{N}{2} (1-P)z +\frac{N}{6} z^3,& 0\leq z \leq 1
       \,           ,
       \\
       \frac{N}{6} -\frac{PN}{2} z +\frac{N}{2}z^2, & 1\leq z \leq (P-1)
       \,           ,
       \\
       -\frac{N}{2} (P-1)(P-z) +\frac{N}{6}  (P-z)^3, & (P-1)\leq z \leq P
       \,           .
     \end{cases}
    \end{equation}
The function $\alpha(z)$ vanishes at $z=0$ and $z=P$, is continuous at $z=1$ as well as at $z=(P-1)$ and its derivative is continuous at those two points. 

We stress that while we will present the details and plots of the numerical analysis for \cref{eq: alpha_exampe1,eq: alpha_exampe2}, we have thoroughly checked many more examples and all lead to the same qualitative behaviour. 

Having described the main characteristics of the quiver theories and how they connect to the supergravity defining function, we proceed to analytically prove that these quivers are exactly non-integrable. The way to do this follows from the simple argument that all quivers start with an $SU(N)$ factor, and this translates to the $\alpha(z)$ assuming a very simple form, namely $\alpha(z)=a_1 z + a_2 z^3$ in the first part of the interval. This can be explicitly seen from the 2 examples we presented above. We have obtained the NVEs in the most general case, without assuming anything on the form of the $\alpha(z)$. We will use the simplest choice $\alpha(z)=a_1 z + a_3 z^3$ to demonstrate that the ordinary quivers are all necessarily non-integrable.

We have explained Kovacic's algorithm sufficiently in \cref{app: kovacic}. Here, we just apply its logic on the NVEs.

With this election of the defining function, the potential appearing in the NVE for $\varrho$ adopts the form
\begin{eqnarray}
\mathcal{V}_\varrho(\tau)=\frac{27a_1a_3E^2\pi^2}{(36a_1\pi^2+a_3E^2\tau^2)^2}+\frac{a_3E^2(m-1)}{4(36a_1\pi^2+a_3E^2\tau^2)}
\end{eqnarray}
This potential fall into Case 3 in the classification given in Kovacic's algorithm. That is, $\mathcal{V}_\varrho(\tau)$ has order 2 at infinity and the order of the poles does not exceed 2. So in principle this NVE could accept Liouvillian solutions. 

The potential appearing in the NVE for $\theta_1$ adopts the form
\begin{equation}
\begin{aligned}
\mathcal{V}_{\theta_1}&=\alpha_1^2+\frac{27a_1\pi^2}{a_3E^2}\frac{1}{\bigg(\frac{36a_1\pi^2}{a_3E^2}+\tau^2\bigg)^2}-\frac{1}{4\bigg(\frac{36a_1\pi^2}{a_3E^2}+\tau^2\bigg)}+\frac{-\alpha_1 a_3\sqrt{m}E^2}{36a_1\pi^2}\sqrt{-\bigg(\frac{36a_1\pi^2}{a_3E^2}+\tau^2\bigg)} \\&-\frac{a_3E^2\sqrt{m}\alpha_1 \tau^2}{36a_1\pi^2}\frac{1}{\sqrt{-\bigg( \frac{36a_1\pi^2}{a_3E^2}+\tau^2\bigg)}}
\end{aligned}
\end{equation}
However, the above expression is not written in terms of rational functions, and hence it is not in an appropriate form for us to apply the criteria of Kovacic's algorithm. As we will see, however, in the later sections where we perform numerical analysis, this does not stop us from determining that the quivers of these type are not classically integrable. We note, that the resulting potential for the $\theta_2$ NVE also cannot be written in terms of rational functions, and hence it is not amenable to the algorithm by Kovacic. 
\subsection{A more exotic and special quiver}\label{sec: exotic_quiv}
We claim that the situation above is drastically different when the quiver instead of being linear is of sinusoidal shape. This means that the defining function, $\alpha(z)$, in the supergravity description is given by: 
\begin{equation}\label{eq: special_sin}
    \alpha(z) \propto  \sin (z)
    \,      ,
\end{equation}
up to some unimportant numerical coefficients.

We feel that some necessary comments are in order: 

\begin{itemize}
    \item This is exactly the solution that led to an integrable $6$-dimensional theory in the mother $\text{AdS}_7$ backgrounds \cite{Filippas:2019puw}. We have already, thoroughly, explained why this situation is by no means a trivial statement of our work. Instead, it presents itself as an interesting case scenario for further and deeper explorations between the sinusoidal-shaped quivers in massive type IIA supergravity and classical integrability. 
    \item It is obvious that for the solution given by \cref{eq: special_sin} we cannot have its derivative -of any order- being equal to zero. We know, however, that $\alpha^{\prime \prime \prime}(z)$ is related to the $0$-form in the Ramond-Ramond sector; the Romans mass. Hence, the behaviour of $\alpha^{\prime \prime \prime}(z)$ is associated with the distribution of D$8$-branes in the background. This resembles closely the interpretation that was given in \cite{Filippas:2019puw} for the integrable quiver in the mother $\text{AdS}_7$ setup. 
    \item This leads us to suggest the following physical interpretation for this special solution of the characteristic function. For $\alpha(z)$ given by \cref{eq: special_sin} we have D$8$-brane sources that are smeared all along the range of the $z$-dimension, instead of them being sharply localised as in the ordinary solutions; the ones presented in \cref{sec: generic_alpha}.
\end{itemize}

Next, we proceed to examine the consequences of choosing \cref{eq: special_sin} as the defining function for our supergravity solution on the various NVEs we have derived in \cref{sec: closed_strings}. We make our definition more precise by specifying $\alpha(z)=A \sin(\omega z)$, with $\omega$ being associated to the number of $D6$-branes. In the above $A \in \mathbb{R}_{\geq 0}$ and $\omega \in \mathbb{R}_{\geq 0}$

For this choice of the defining function $\alpha(z)$ the geometry in \cref{eq: geometry} becomes:
\begin{equation}\label{eq: geometry_special}
\begin{aligned}
ds^2 = 
&12 \sqrt{\frac{6}{m}} \frac{\pi}{\omega} ds^2_{\text{AdS}_5} + \sqrt{\frac{6}{m}} \frac{\pi}{\omega} ds^2_{\mathbb{H}^2} + \sqrt{\frac{3m}{2}} \pi \omega dz^2 + 
\\
&\sqrt{6m} \frac{\pi}{\omega} \frac{1}{3m+2 \cot^2(\omega z)}  \left( ds^2_{S^2} + \cosh\theta_1 \sin^2 \theta_2 ds^2_2 \right)		\,			.
\end{aligned}
\end{equation}
It should be obvious from the form of \cref{eq: geometry_special} that up to the fibration, denoted by $ds^2_2$, the geometry has reduced to that of integrable submanifolds. We note that by integrable submanifolds we mean that the Polyakov action on $\text{AdS}_n$, $\text{dS}_n$, $\text{S}^n$ and $\mathbb{H}^n$ is known to be integrable in the absence of a $B_2$-field; see \cite[appendix B]{Filippas:2019puw} for the construction of the Lax connection in these cases. 

Note that a similar, but indeed much easier to understand, situation arose in the study of the mother $\text{AdS}_7$ backgrounds \cite{Filippas:2019puw}. Indeed, when studying the mother theory there is, of course, no such complication as a non-trivial fibration. This is why in the UV, where the theory flows to the $6$-dimensional $\mathcal{N}=(1,0)$ SCFT it was more straightforward to obtain a Lax.

This is our first piece of evidence, although not strong and conclusive, and is entirely based on the isometries that are present in the setup. It is, also, once again quite obvious why the status of integrability in the $\text{AdS}_5$ does not immediately follow from the corresponding statement of the mother $\text{AdS}_7$ solution. 

We proceed to quickly inspect the NVEs for the $\rho$ and $\theta_1$ in this case and offer some comments for the NVE of $\theta_2$ that, also, connect to a similar peculiarity in the $\text{AdS}_7$ backgrounds. 
	\subsection*{NVE for \texorpdfstring{$\rho$}{rho}}
In this case, \cref{eq: NVErho}, becomes:
\begin{equation}
\varrho''(\tau)+\frac{E^2m \omega^2}{864\pi^2}\varrho(\tau)=0    
\end{equation}
which has the simple solution:
\begin{equation}
\varrho = \cos\left(\frac{E}{12} \sqrt{\frac{m}{6}} \frac{\omega}{\pi} \tau\right) + \sin\left(\frac{E}{12} \sqrt{\frac{m}{6}} \frac{\omega}{\pi} \tau\right)
\,          ,
\end{equation}
that is clearly Liouville integrable. 
	\subsection*{NVE for \texorpdfstring{$\theta_1$}{theta1}}
\Cref{eq: NVEtheta1} becomes:
\begin{equation}
\vartheta_1^{\prime \prime}(\tau) + \frac{1}{6} \alpha_1 (E-6 \alpha_1) \vartheta_1(\tau) = 0
\,          ,
\end{equation}
which admits the Liouville integrable solution:
\begin{equation}
\vartheta_1 = e^{\sqrt{-\alpha^2_1 + \frac{\alpha_1 E}{6}} \tau} + e^{-\sqrt{-\alpha^2_1 + \frac{\alpha_1 E}{6}} \tau}
\,          .
\end{equation}
	\subsection*{NVE for \texorpdfstring{$\theta_2$}{theta2}}
The NVE of $\theta_2$ is much more cumbersome to analyse, even for the case of $\alpha(z)=A \sin(\omega z)$. This has to do with the increased complexity in the equations of motion of that particular coordinate. Similar difficulties appeared, also, in the analysis of the $\text{AdS}_7$ backgrounds where we know the sinusoidal quiver is integrable, since we have an explicit Lax connection \cite{Filippas:2019puw}. Truly, this was observed in their context for the NVE of the angular $\chi$-dimension and we refer the reader to \cite[equations (A.10) and (A.11)]{Filippas:2019puw} for the details in that setup. What we face here is entirely equivalent to that situation. From an inspection of the NVEs presented in \cref{sec: closed_strings} we can see that in the NVE for the $\theta_2$ coordinate, and more specifically its Schr\"odinger form, the warp factors $f_4$ and $f_6$ appear, unlike in the other two cases that only the $f_1$ and $f_2$ are present. This, in turn, results in obtaining a very complicated sum of large fractions, from which it is not possible to make any statements on the status of integrability for the string soliton. We will omit any further discussion here and we will see from the numerical analysis that this is not a problem, as we are not led to chaotic behaviour when studying this NVE. Borrowing intuition from the $\text{AdS}_7$ results, see \cite{Filippas:2019puw}, we provide the hand-wavy argument that this is an unimportant complication that should not be discouraging at all. 
\section{Numerical analysis of strings and their dynamics}\label{sec: numerics}
In this section, we proceed to carry out an exhaustive numerical analysis which will allow us to elaborate more on the analytic results obtained in the previous section. The numerical analysis, also, provides us with a more solid understanding of all four-dimensional $\mathcal{N}=1$ quiver theories that are associated with these massive type IIA vacua. We will be able to compare the dynamics of any generic non-integrable and chaotic quiver to those of the special ones that are integrable. As we will see, deforming the special solution that we have found exhibiting integrability even by some small amount results to signatures of chaos. This complements our previous findings which were based on the study of the NVEs. 

We perform numerical computations by studying the dynamics of classical strings that are propagating in the backgrounds we consider in this work. These considerations can be used to demonstrate that the dynamics in the phase-space of these classical strings exhibits characteristic signatures of chaos, and are thus non-integrable.

It is important at this point to pause for a brief moment and make a comment, once more, pertaining to the relation of chaos to the statement of a system's integrability. It is well understood that the non-integrability of a system does not, necessarily, imply chaos. However, on the other hand, chaotic dynamics is indicative of the absence of integrability. It  is, also, worthwhile mentioning that in the context of the gauge/gravity duality  all known examples hitherto that have been argued to be non-integrable were also shown to be chaotic; see also \cite{Nunez:2018ags} where these ideas were explained. 

We describe the general idea in order to explain our approach. We are following \cite{Ott:2002book}\footnote{See also the related numerical analysis in \cite{Nunez:2018ags} for propagating strings in the mother $\text{AdS}_7$ backgrounds, and also \cite{Nunez:2018qcj} for a similar analysis in the Gaiotto-Maldacena backgrounds.} in our discussion.

Let us consider a dynamical system. The time-evolution of this system is determined by a set of differential equations. This allows us to calculate the state of a system at a time, which we denote by $t_{\text{final}}$, knowing an earlier state of the system at some fixed point in time, which we denote by $t_{\text{earlier}}$. Following this logic we can reach the beginning of the dynamical time evolution of the system at $t_0$. Therefore, knowing the state of the system at the initial time $t_0$ we can, in principle, solve the system at hand at any time. 

Such a dynamical system is considered to be chaotic if it is sensitive to its initial conditions. Namely, if the nearby points of the system depart from one another in an exponential way along the time flow. This is making it impossible to accurately predict its dynamical behaviour based on the states of previous times. A simple, but illustrative example of this is as follows: let us assume that we have two adjacent initial conditions describing points that lie  arbitrarily close to each other. We use $h_1(t_0)$ and $h_2(t_0)= h_1(t_0) + \epsilon$ to denote these points and $\epsilon$ is a small parameter. We declare that the system exhibits chaotic dynamics when $|h_1(t) - h_2(t)| \sim e^{\lambda t}$. This is under the assumption that the system's trajectories in the phase-space are bounded. This criterion on the boundedness of the trajectories is essential in the sense that it rules out a possible situation in which the trajectories are moving towards infinity. In that scenario, the exponential divergence is because they are moving apart \cite{Ott:2002book}. This is the so-called trivial case \cite{Nunez:2018ags}. 
 
The case of our interest is that of the motion of classical strings that are positioned at the centre of the $\text{AdS}_5$ spacetime, while moving and rotating in the internal space of the ten-dimensional solution. This is described by the system of equations that we derived in \cref{eq: el_eqns,eq: stresstensor} or their analogous ones, see \cref{eq: conjugate_momenta,eq: hamiltonian} for the Hamiltonian formulation. The coordinates of the internal space are bounded and the same holds true for the associated conjugate momenta owing to the conservation of the  Hamiltonian.

We will make a specific choice for the linear quiver to be used in the sections below as an illustrative example of our methods. We have already presented two concrete and interesting examples of quivers taken from \cite{Merrikin:2022yho}. We will make another concrete choice that will facilitate our analysis. This is given by:
\begin{equation}\label{eq: linear_quiver_1}
    \alpha(z) = - 81 \pi^2 5 \left( -6z+\tfrac{1}{6}z^3 \right)
    \,         .
\end{equation}

Note that a similar solution was studied in \cite{Nunez:2018ags} for the seed $\text{AdS}_7$ solutions. The interest in this is because it presents itself as the simplest, non-trivial solution with good behaviour. 
\subsection{The dynamical evolution of the system and power spectra}
\begin{figure}[ht!]
\centering
\includegraphics[width=1\linewidth]{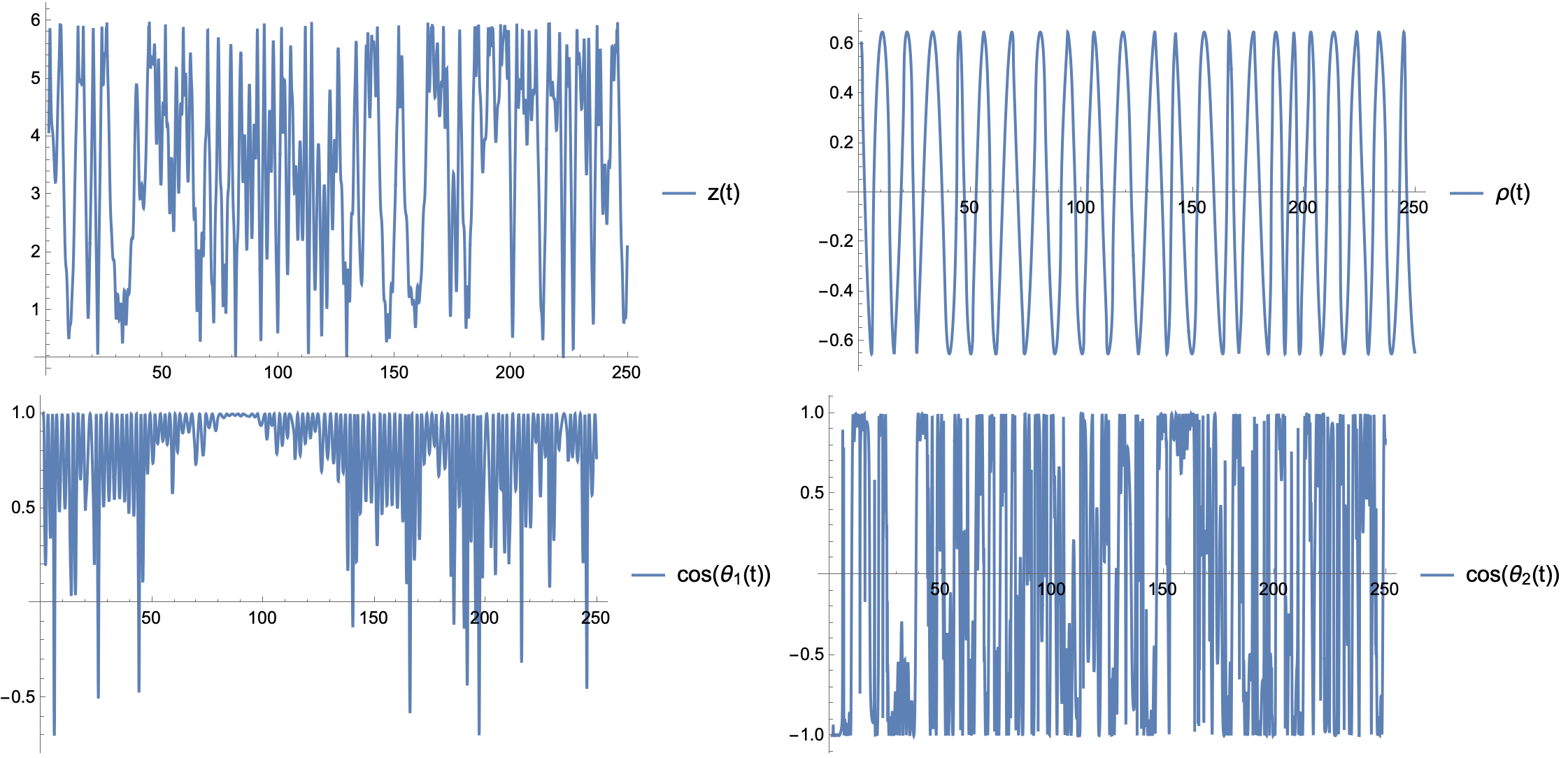}
   \caption{The dynamical evolution of the various string coordinates in time. The parameters chosen are: $\alpha_1=\alpha_2=1$, $m=1$, and $\Psi_0=0$.}
\label{fig: dynamics_linear_quiver_1}
\end{figure}

We begin our numerical analysis section with the study of the classical string dynamics. This is the simple statement that we allow for time to evolve and we solve the differential equations resulting from the Lagrangian density numerically, for the choice of a linear quiver corresponding to \cref{eq: linear_quiver_1}. Subsequently, we plot the string motion along the various coordinates against time. This is shown in \cref{fig: dynamics_linear_quiver_1}.

\begin{figure}[ht!]
\centering
\includegraphics[width=1\linewidth,height=115pt]{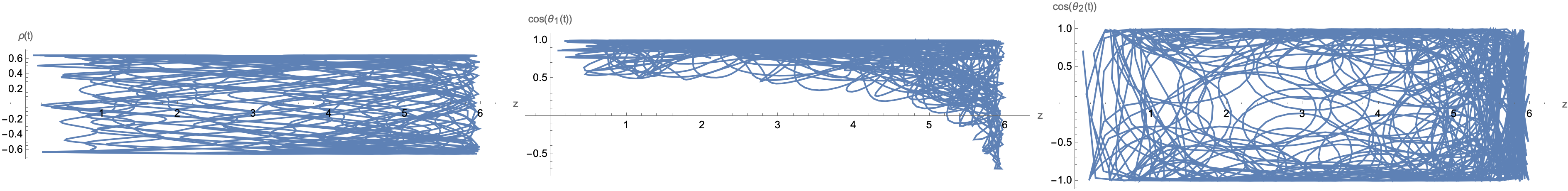}
   \caption{The various frozen trajectories in the $(z(t),\texttt{any coordinate})$-plane. The parameters chosen are: $\alpha_1=\alpha_2=1$, $m=1$, and $\Psi_0=0$.}
\label{fig: fromzplane}
\end{figure}

It should be quite obvious that, except for the $\rho$-coordinate, the string motion along the rest is chaotic. It is, perhaps, quite instructive to examine the trajectories in the relative plane of coordinates for the string motion. We start by depicting the dynamical evolution in the $(z(t),\texttt{any coordinate})$-plane in \cref{fig: fromzplane}

\begin{figure}[ht!]
\centering
\includegraphics[width=1\linewidth,height=200pt]{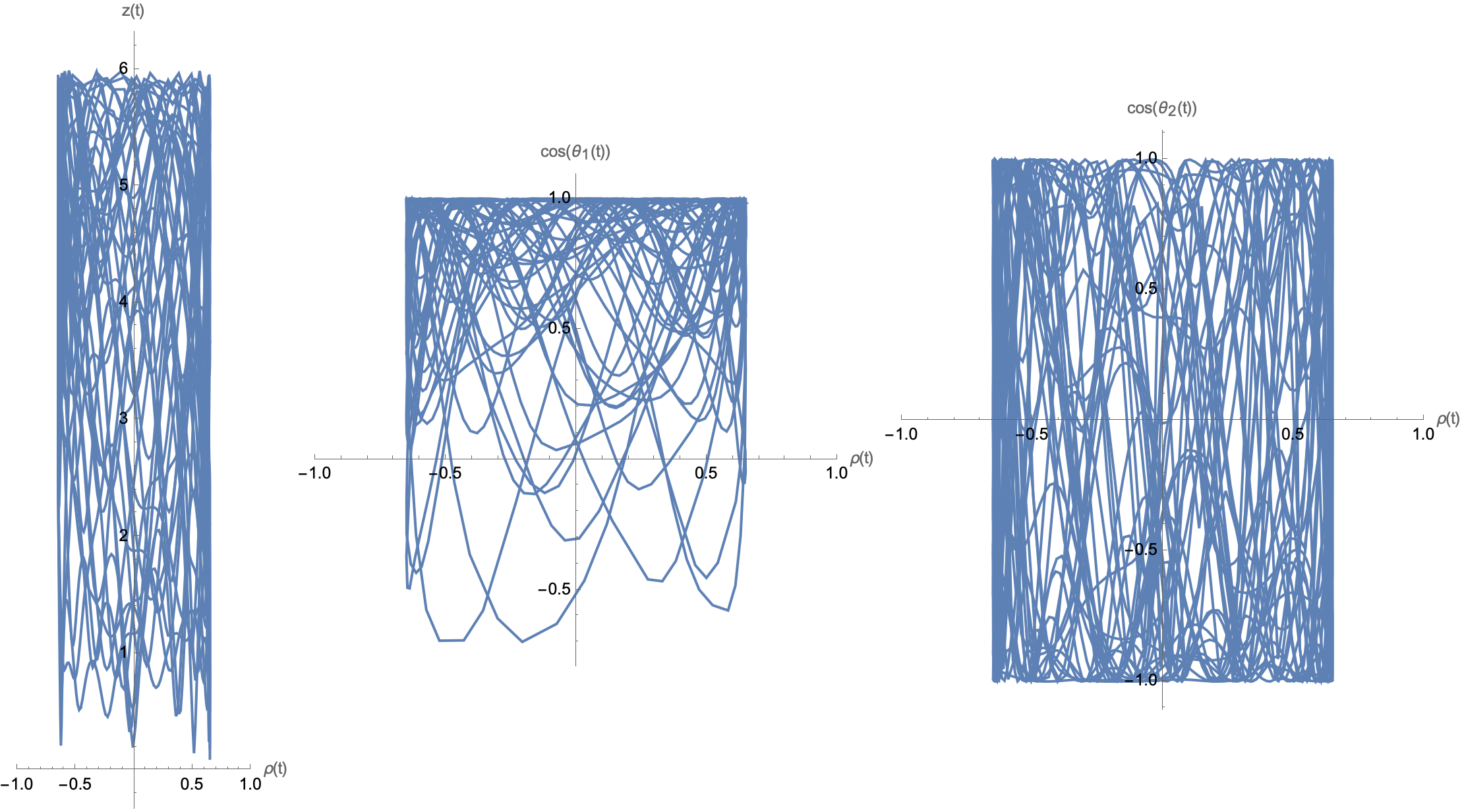}
   \caption{The various frozen trajectories in the $(\rho(t),\texttt{any coordinate})$-plane. The parameters chosen are: $\alpha_1=\alpha_2=1$, $m=1$, and $\Psi_0=0$.}
\label{fig: fromrhoplane}
\end{figure}

\begin{figure}[ht!]
\centering
\includegraphics[width=1\linewidth]{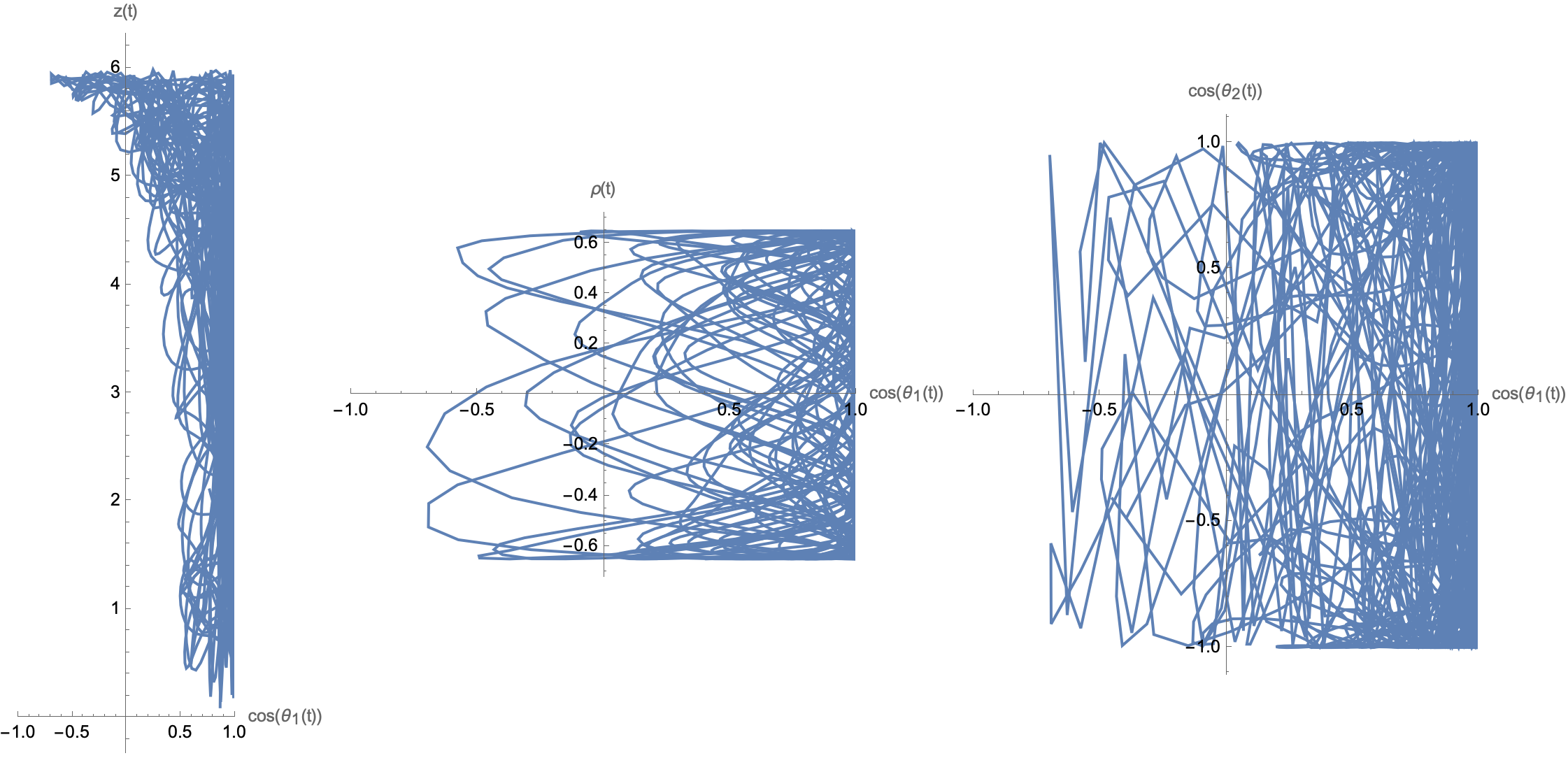}
   \caption{The various fronzen trajectories in the $(\rho(t),\texttt{any coordinate})$-plane. The parameters chosen are: $\alpha_1=\alpha_2=1$, $m=1$, and $\Psi_0=0$.}
\label{fig: fromtheta1plane}
\end{figure}

\begin{figure}[ht!]
\centering
\includegraphics[width=1\linewidth]{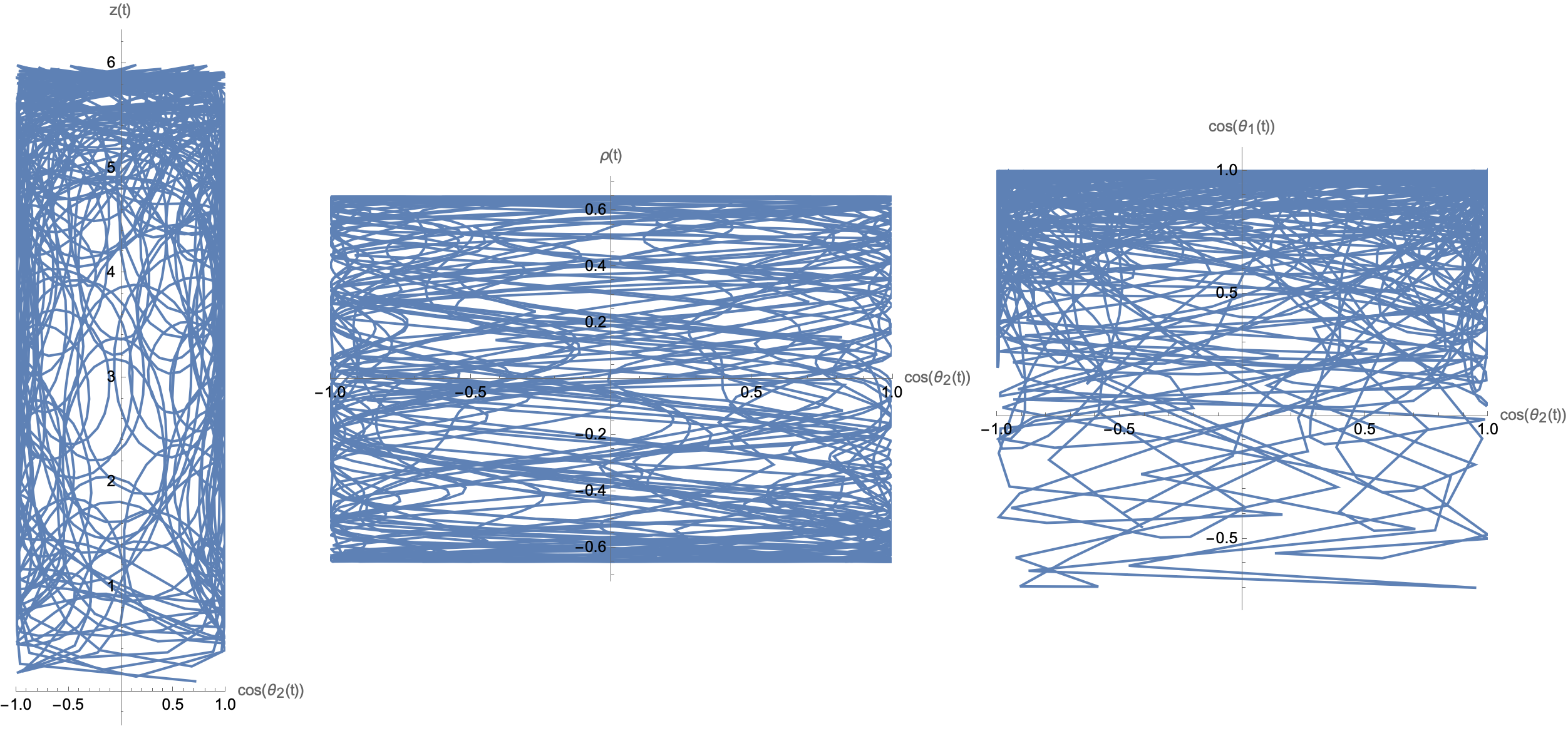}
   \caption{The various frozen trajectories in the $(\rho(t),\texttt{any coordinate})$-plane. The parameters chosen are: $\alpha_1=\alpha_2=1$, $m=1$, and $\Psi_0=0$.}
\label{fig: fromtheta2plane}
\end{figure}

A very quick inspection of \cref{fig: fromzplane} reveals that the string-dynamics are highly disorganized and lack any integrability. More precisely what we find out is that the motion is not periodic in any of those planes, and the path of the string does not close on itself. We will see a confirmation of this picture and intuition in the corresponding power spectrum, as well. We point out that one can choose one of those planes, e.g the $(z,\rho)$-plane, and allow the string to be energetically excited. And then proceed with this to the rest of the planes. The same qualitative conclusions can be reached, of course. We refrain from doing so, to avoid providing too many plots, and instead we opt to provide the other relative planes for a particularly frozen moment, for completeness. These are depicted for the $(\rho(t),\texttt{any coordinate})$-plane in \cref{fig: fromrhoplane}, and for $(\theta_1(t),\texttt{any coordinate})$- and $(\theta_2(t),\texttt{any coordinate})$-planes in \cref{fig: fromtheta1plane,fig: fromtheta2plane} respectively. 

\begin{figure}[ht!]
\centering
\includegraphics[width=1\linewidth]{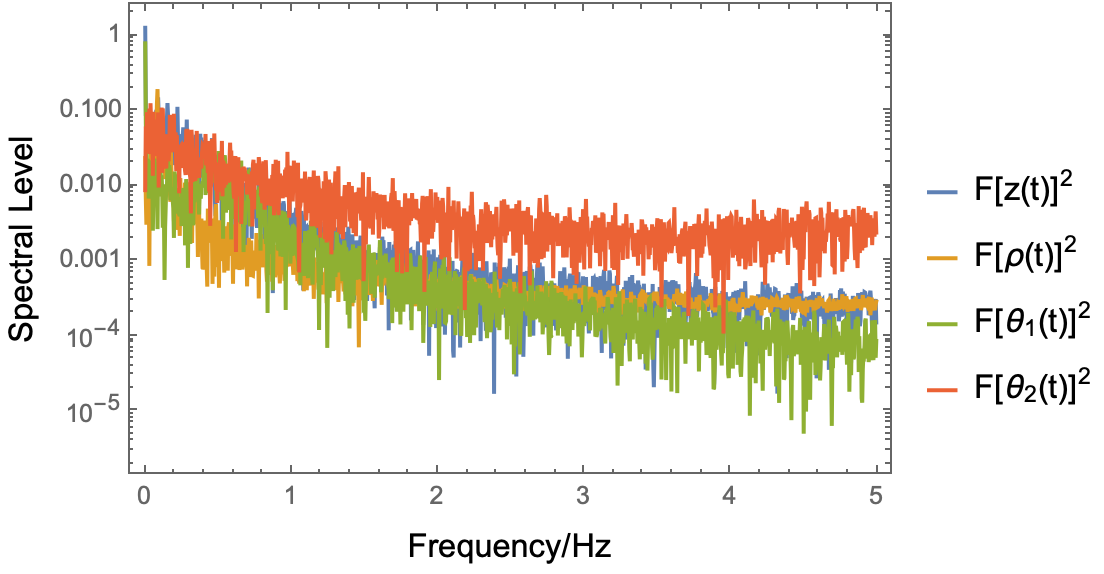}
   \caption{The power spectra for the ($z(t),\rho(t),\theta_1(t),\theta_1(t)$) trajectories. The parameters chosen are: $\alpha_1=\alpha_2=1$, $m=1$, and $\Psi_0=0$.}
\label{fig: powerspectrumlinearquiver}
\end{figure}

We now move on to discuss the corresponding power spectra following \cite{Ott:2002book,Nunez:2018ags,Nunez:2018qcj}. We start by taking the Fourier transform of the numerical evolution of string motion, depicted in the plots \cref{fig: fromzplane,fig: fromrhoplane,fig: fromtheta1plane,fig: fromtheta2plane}. Doing so, we can determine if the corresponding coordinates are periodic, quasi-periodic or chaotic. If a signal has a well-defined period characterized by a frequency $\omega$, the relevant spectrum will have a vertical line at the characteristic frequency of the system. Of course, one has to be careful with subtleties in the numerical errors and estimates when examining these behaviours \cite{Ott:2002book}. As we see, however, in \cref{fig: powerspectrumlinearquiver} the situation is very clear and suggestive. 

As we see from the plot of the power spectra, which is presented in \cref{fig: powerspectrumlinearquiver}, for the various trajectories, there seems to be a fundamental frequency of value 0.01, however, the chaotic behaviour observed in \cref{fig: dynamics_linear_quiver_1} is evident here as well. Truly, very quickly as we increase the frequency, we observe that all the higher harmonics are lost and what is happening is that there is a broad band of noise that is overtaking the spectra. By the end, the full picture is just dominated by pure noise. As we have mentioned previously in the examination of the dynamics in the relative planes and not in Fourier space, one can focus on a particular plane -or even choose all of them- and proceed to fine tune the initial conditions of the system in such a way that the string is getting energetically excited. Then, proceed to examine the quantitative behaviour of the system. While this is an interesting study it presents itself as immaterial for our purposes, since we would reach the same qualitative behaviours. Since in \cref{fig: powerspectrumlinearquiver} there is no sign to indicate that higher harmonics could, perhaps, survive and compete with the noise that is manifested, there is no need to further delineate this study and we can make a conclusive argument even at this point. 

\begin{figure}[ht!]
\centering
\includegraphics[width=1\linewidth]{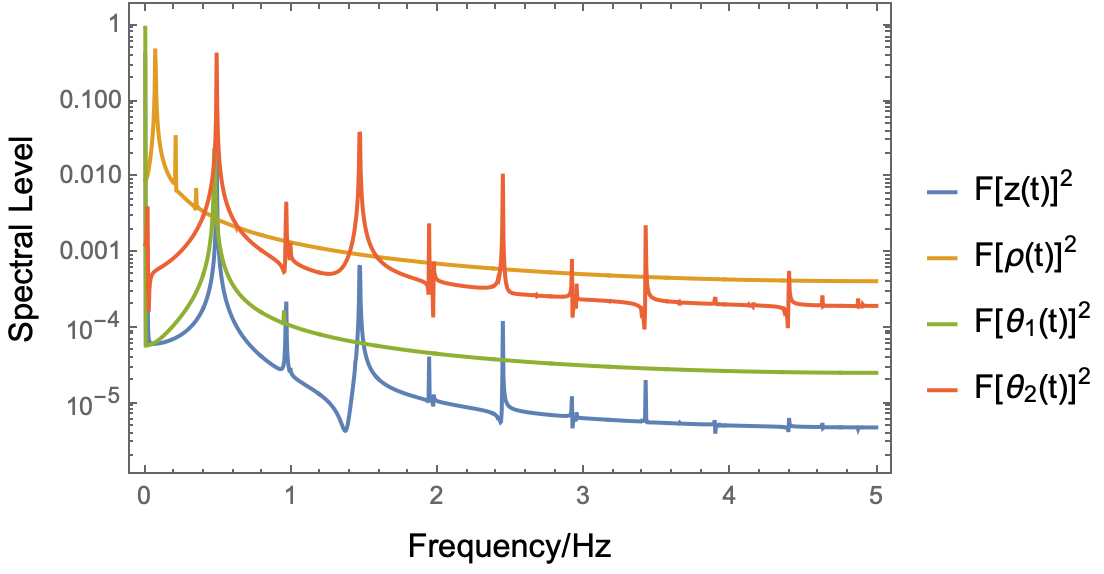}
   \caption{The power spectra for the ($z(t),\rho(t),\theta_1(t),\theta_1(t)$) trajectories. The parameters chosen are: $\alpha_1=\alpha_2=1$, $m=1$, and $\Psi_0=0$.}
\label{fig: powerspectrumsin}
\end{figure}

Now, we turn our attention very quickly to the special quiver that corresponds to a sinusoidal shape. For the choice $\alpha(z) = A \sin (\omega z)$ we repeat the same analysis that we have presented above. We refrain from providing all the plots and we present our findings for the power spectrum in this case since we can derive conclusive evidence from that. As we can observe by inspecting \cref{fig: powerspectrumsin} the situation is drastically different compared to \cref{fig: powerspectrumlinearquiver}.

It is clear that in \cref{fig: powerspectrumsin} none of the trajectories is losing its higher harmonics as we increase the frequency, and there is no sign of noise trying to overpower the spectrum. We have checked that even if we energetically excite the string and repeat the same analysis, we can reach the same qualitative output. This is another strong indicator that the choice of a sinusoidal quiver is very special with regards to the status of integrability. 

The observed behaviour in the \cref{fig: powerspectrumsin} is also manifested in the dynamical evolution of the string motion in position-space, as well as from the relative planes. Truly, if one plots the dynamical evolution of the string along the various dimensions against time, then one sees clearly a periodic and canonical motion, rather than the erratic and chaotic behaviour observed in \cref{fig: dynamics_linear_quiver_1}. The corresponding statement from the relative planes of motion, is that the trajectories all close on themselves and form figures with clear patterns. 

\begin{figure}[hb!]
\centering
\begin{subfigure}{.55\textwidth}
  \centering
  \includegraphics[width=1\linewidth]{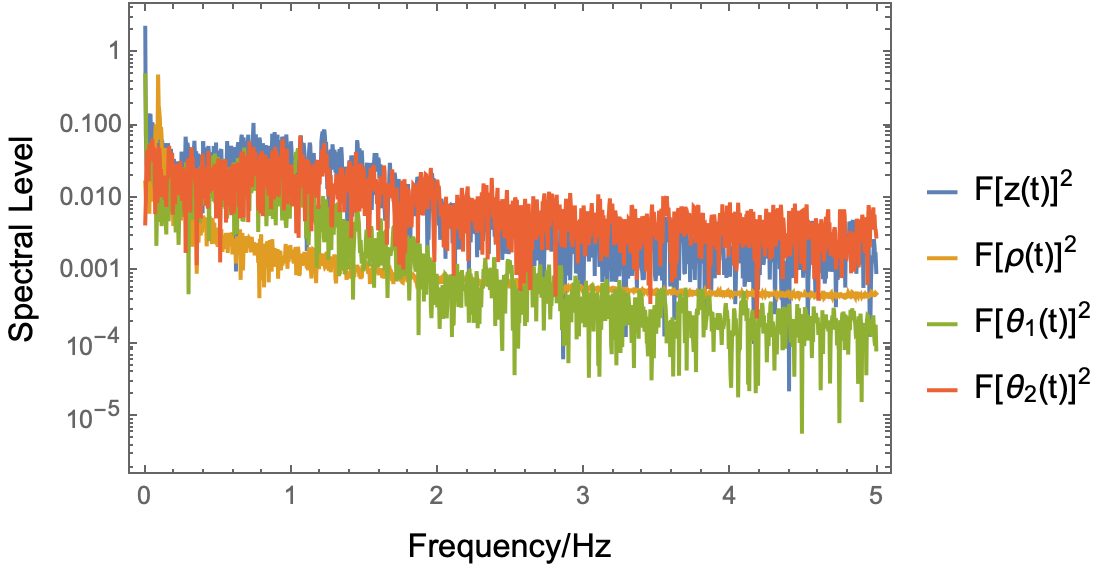}
  \caption{The power spectrum for \cref{eq: alpha_exampe1}.}
  \label{fig:sub1ps}
\end{subfigure}%
\begin{subfigure}{.55\textwidth}
  \centering
  \includegraphics[width=1\linewidth]{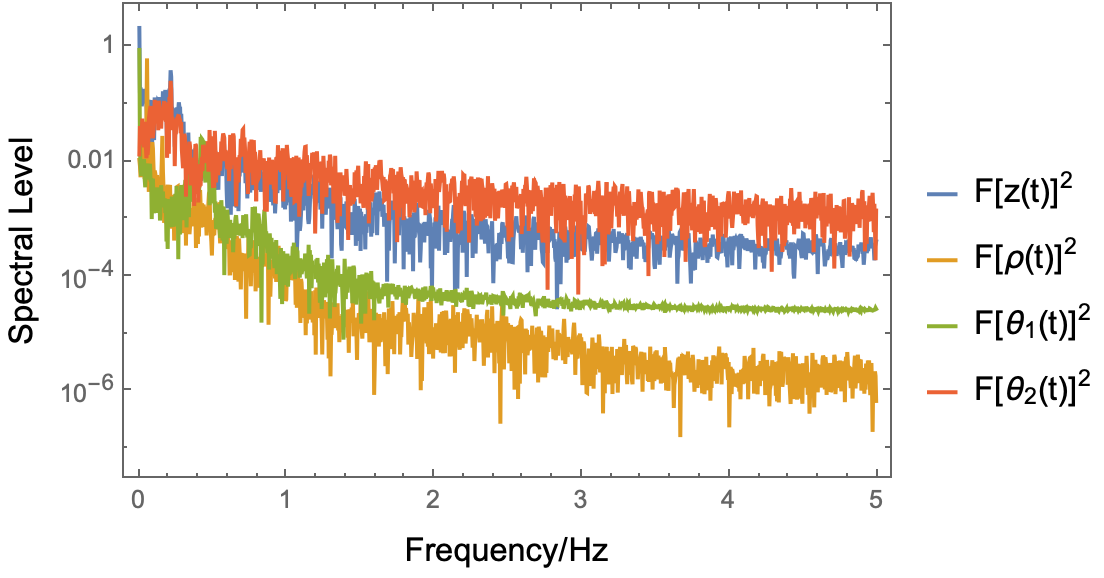}
  \caption{The power spectrum for \cref{eq: alpha_exampe2}.}
  \label{fig:sub2ps}
\end{subfigure}
\caption{Power spectra for the two quivers discussed in \cref{eq: alpha_exampe1,eq: alpha_exampe2}}
\label{fig: powerspectra_examples}
\end{figure}

Before closing this subsection, we present the results from the computations of the power spectra for the two examples that we described in \cref{eq: alpha_exampe1,eq: alpha_exampe2}. These results are shown in \cref{fig: powerspectra_examples}.

	\subsection{The spectrum of Lyapunov exponents}
The spectrum of Lyapunov exponents is considered to be a quantity determining that the dynamical information is lost when a system exhibits chaos. The loss of information is sourcing the dynamical Kolmogorov-Sinai entropy in a chaotic setup. When we consider a given dynamical system that is described by a non-zero Lyapunov exponent, the explanation that is associated with this phenomenon is that the time evolution of two trajectories that are close to each other in the phase-space turns out to be extremely sensitive to an infinitesimal change in the initial conditions. This effect is amplified exponentially at sufficiently large times. More simply put, the existence of a non-zero Lyapunov exponent, for a point $Y=(q,p)$ in the phase space with initial condition $Y_0=(q_0,p_0)$ at $t=0$ is given by:
\begin{equation}
\lambda = \lim_{\substack{t \rightarrow \infty\\ \Delta Y_0 \rightarrow 0}}\frac{1}{t}\log \frac{\Delta Y (Y_{0}, t)}{\Delta Y (Y_0 ,0)}		\,			,
\end{equation}
and is intimately related to the degree of randomness associated with the dynamical phase space of a Hamiltonian system \cite{Rigatos:2020hlq,Nunez:2018ags,Nunez:2018qcj}.

It is  measures the rate of separation between two trajectories that lie arbitrarily close in the phase-space of the theory. The function $\Delta Y (Y_0 , t)$ measures the separation between two such trajectories as a function of this initial location. For chaotic systems, we end up obtaining
\begin{equation}
\Delta Y (Y_{0}, t) \sim \Delta Y (Y_{0}, 0)e^{\lambda t}.
\end{equation}

We proceed by providing the details for the computation and implementation of the spectrum of Lyapunov exponents. In the description of the spectrum we use $\lambda_i$ to denote the $i$-th Lyapunov Charcteristic Exponent, henceforth abbreviated LCE, according to the standard nomenclature in the literature. We follow \cite{sandri1996} in all the details and numerical tecnhiques of our implementation. This has been used in many case in the study of chaos in holographic setups, see for example \cite{Nunez:2018ags,Nunez:2018qcj}. We use the notational conventions of \cite{Nunez:2018ags,Nunez:2018qcj} for convenience. 

For the computation of the Lyapunov exponents we choose the initial conditions in such a way as to satisfy the Hamiltonian constraint, i.e $\mathcal{H}=0 $. This in turn implies that for a $2N$ dimensional phase space,
\begin{equation}
\sum_{i=1}^{2N}\lambda_{i}=0
\end{equation}
This is what measures the exponential growth associated with the $i$-th direction in the phase-space. For some systems, the sum of all positive Lyapunov exponents measures the Kolmogorov-Sinai entropy production during the dynamic evolution in the phase-space.

We substitute appropriate initial conditions into the dynamical equations in order to generate a solution. Upon choosing these initial conditions for the phase-space variables and make sure that they satisfy the vanishing of the Hamiltonian, we find the corresponding Lyapunov spectrum for each of the quiver configurations. Notice that, in our analysis, we are eventually computing four LCEs ($\lambda_{i}$) that characterise a four-dimensional dynamical phase space. 

The prescription of \cite{sandri1996} is the following:

We consider a generic $2n$-dimensional smooth dynamical canonical system, which can be written as:
\begin{equation}\label{dynsys}
\dot{q}=\mathfrak{V}(q)
\,      ,
\end{equation}
in all generality, and where $q(t)$ is the $2n$-dimensional state vector $q=\big(\vec{Y}(t), \vec{P}(t) \big)$ at time $t$, $\dot{q}=\frac{dq}{dt}$ and $\mathfrak{V}$ is a vector field defined on an open set $U$ of the phase-space manifold. This vector field generates a flow $f$ given by:
\begin{equation}
\dot{f}^t(q)=\mathfrak{V}(f^t(q)) \ \ \  \mathrm{for \ all} \ q \in U,t \in \mathrm{R}
\,          ,
\end{equation} 
with $f^t(q)=f(q,t)$.

Let us follow the evolution under the flow of two arbitrarily close points in phase-space, $q_0$ and $q_0+\delta_0$, where the subindex 0 denotes the initial time, with $\delta_0$ being a small deviation of $q_0$. After a time $t$, the deviation $\delta_t$ evolves to:
\begin{equation}\label{tangent}
\delta_t \equiv f^t(q_0+\delta_0)-f^t(q_0) \approx D_{q_0} f^t(q_0) \cdot \delta_0
\,              ,
\end{equation}
where the difference in trajectories has been approximated with the gradient $D_{q_0}f^t(q_0)$. We can compute the averaged exponential growth of two trajectories, as
\begin{equation}\label{LCE}
\lambda(q_0,\delta_0)=\lim_{t \rightarrow \infty} \frac{1}{t} \log \frac{||\delta_t||}{||\delta_0||}=\lim_{t \rightarrow \infty} \frac{1}{t} \log ||D_{q_0} f^t(q_0) \cdot \delta_0||
\,          ,
\end{equation} 
where $||\delta||$ denotes the norm of the vector $\delta$ according to standard nomenclature; e.g \cite{sandri1996,Oseledec1968}. In the case that $\lambda>0$, the nearby orbits will exponentially diverge. By making some mild assumptions on the vector field, the limit, given by \cref{LCE}, exists. In this case, the limit is finite and defines the largest LCE $\lambda_1$.

We can introduce the LCEs of order $p$, with $1 \leq p \leq n$, as a means to describe the mean rate of growth of a $p$-dimensional volume in the tangent-space. By constructing a solid of which each face is a parallelogram $U_0$ in the tangent space and with edges given by $\delta_1,...,\delta_p$, we define LCEs of order $p$ as:
\begin{equation}\label{volp}
\lambda^p(q_0,U_0)=\lim_{t \rightarrow \infty} \frac{1}{t} \log [\mathrm{vol}^p (D_{q_0} f^t(U_0))]
\,          ,
\end{equation}
where $\mathrm{vol}^p$ is the canonical volume-form.

There exist $p$ linearly independent vectors $\delta_1,...,\delta_p$ satisfying\cite{Oseledec1968}:
\begin{equation}\label{LCEp}
\lambda^p(q_0,U_0)= \sum_{k=1}^p\lambda_k
\,              .
\end{equation}
$\delta_t$ as defined in \cref{tangent} dynamically evolves as:
\begin{equation}
\dot{\Psi}_t(q_0)=D_q \mathfrak{V}(f^t(q_0)) \cdot \Psi_t(q_0), \ \ \ \Psi_0(q_0)=\mathbb{1}
\label{fullsys}
\end{equation}
where we have denoted $\Psi_t(q_0)=D_{q_0} f^t(q_0)$, and $\mathbb{1}$ denotes the identity. The trajectories in phase-space are obtained by integrating the following system:
\begin{align}
 \Bigg \{\begin{array}{ccr}
\dot{q}  \\
\dot{\Psi}  
\end{array} \Bigg \}= \Bigg \{\begin{array}{ccr}
\mathfrak{V}(q) \\
D_q \mathfrak{V}(q) \cdot \Psi
\end{array} \Bigg \},  \ \ \  \Bigg \{\begin{array}{ccr}
q(t)  \\
\Psi(t_0)  
\end{array} \Bigg \}= \Bigg \{\begin{array}{ccr}
q_0 \\
\mathbb{1}
\end{array} \Bigg \}
\label{fullsysarray}
\end{align}

The spectrum of LCEs, will be computed following the algorithmic process described in \cite{Benettin1980}. This procedure amounts to the computation of the order-$p$ LCEs as defined in equation \cref{LCEp}. The main upshot of the algorithm is a repeated application of the Gram-Schmidt procedure which simplifies the implementation \cite{Oseledec1968}. We provide the main steps, very briefly, below\footnote{See also \cite{Nunez:2018ags} for an exposition on this approach.}: 

From the original set $\{\delta_i \}$, we can construct an orthonormal set $\{\hat{\delta}_i \}$ using the Gram-Schmidt recipe. The $U_0$-shaped volume spanned by ${\delta_1, ..., \delta_p}$ is now constructed as:
\begin{equation}
\mathrm{vol} \{\delta_1, ..., \delta_p \}= ||\hat{\delta}_1||...||\hat{\delta}_p||
\,              .
\end{equation}

The starting point of the procedure is to pick a set of initial conditions $q_0$ together with a matrix of $n \times n$ dimensionality $\Upsilon=\{\delta^0_1,...,\delta^0_n\}$. Applying the Gram-Schmidt procedure we obtain the basis $\hat{\Upsilon}_0=\{\hat{\delta}^0_1,...,\hat{\delta}^0_n\}$. Subsequently, we integrate the \cref{fullsysarray} from $\{q_0,\Upsilon_0 \}$ in a small neighborhood $\mathfrak{T}$, to obtain $q_1=f^{\mathfrak{T}}(q_0)$ and 
\begin{equation}
\Upsilon_1 \equiv [\delta^1_1,...,\delta^1_n]=D_{q_0} f^{\mathfrak{T}}(\Upsilon_0)= \Psi_{\mathfrak{T}}(q_0) \cdot [\delta^0_1,...,\delta^0_n]
\,          .
\end{equation}
The algorithmic process is such that it is repeating the process described above $\mathcal{M}$ times. During the $m$-th step, the $p$-dimensional volume $\mathrm{vol}^p$ defined in \cref{volp} is increased by a factor of $||\mathfrak{a}^m_1||...||\mathfrak{a}^m_p||$, with $\{ \mathfrak{a}^m_1,...,\mathfrak{a}^m_p \}$ being the set of orthonormal vectors calculated from $U_m$. Then:
\begin{equation}
\lambda^p(q_0, \Upsilon_0)=\lim_{m \rightarrow \infty} \frac{1}{m \mathfrak{T}} \sum^{m}_{i=1} \log(||\hat{\delta}^i_1||...||\hat{\delta}^i_p||)
\,             .
\end{equation} 
We immediately obtain:
\begin{equation}
\lambda_p=\lim_{k \rightarrow \infty} \frac{1}{m \mathfrak{T}} \sum^{m}_{i=1} \log ||\hat{\delta}^i_p||
\,             .
\end{equation}
The full Lyapunov spectra is completed by computing the following:
\begin{equation}
\frac{1}{\mathcal{M} \mathfrak{T}} \sum^{\mathcal{M}}_{i=1} \log ||\hat{\delta}^i_1|| \approx \lambda_1,...,\frac{1}{\mathcal{M} \mathfrak{T}} \sum^{\mathcal{M}}_{i=1} \log ||\hat{\delta}^n_1|| \approx \lambda_n 
\,              ,
\end{equation}
for an appropriately chosen $\mathfrak{T}$, until we find them converging.

\begin{figure}[ht!]
\centering
\begin{subfigure}{0.51\textwidth}
  \centering
  \includegraphics[width=1\linewidth]{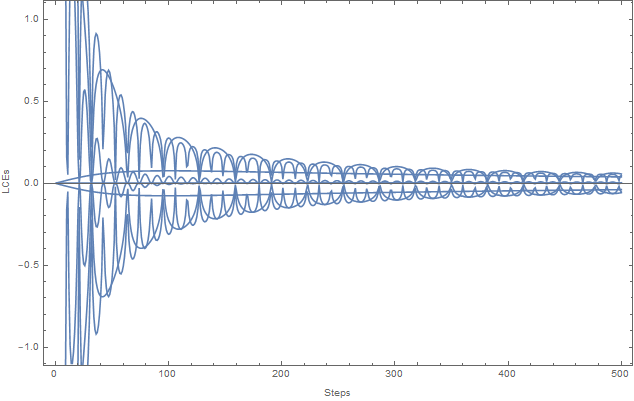}
  \caption{The mass parameter here is: $m = \tfrac{1}{2}$ and the choice for the $\alpha(z) = \sin(z)$.}
  \label{fig:sub1}
\end{subfigure}%
\begin{subfigure}{0.51\textwidth}
  \centering
  \includegraphics[width=1\linewidth]{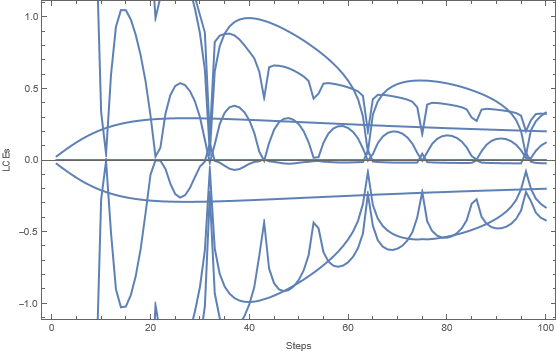}
  \caption{The mass parameter here is: $m = \tfrac{1}{4}$ and the choice for the $\alpha(z) = \alpha(z) = - 81 \pi^2 5 \left( -6z+\tfrac{1}{6}z^3 \right)$.}
  \label{fig:sub2}
\end{subfigure}
\caption{Lyapunov for the integrable and non-integrable quivers}
\label{fig: integrable_lyapunov}
\end{figure}

We plot the LCEs for two examples. Following the discussion above it should be clear that there no chaotic indicators for the choice $\alpha(z) = A \sin(\omega z)$, while for the simple quiver given by \cref{eq: linear_quiver_1} the motion is chaotic and thus there is no chance of Liouville integrability. We present our computations in \cref{fig: integrable_lyapunov}

	\subsection{Poincar\'e sections}
Let us start by considering an $\nu$-dimensional integrable system. The system has $\nu$ independent integrals of motion in involution; namely that the Poisson bracket of any two of these conserved quantities is zero. This in turn implies that the corresponding phase-space trajectories are confined to the surface of an $\nu$-dimensional KAM torus \cite{Ott:2002book}. Upon a change of variables $(q_i, p_i)$ the so-called action-angle variables $(\psi_i, G_i)$, in such a way that the Hamiltonian describing the system depends on $G_i$ only, the related classical paths on this KAM torus are specified entirely in terms of the $\nu$ frequencies, denoted by $(\omega_i)$, which give the associated velocities along each of these angles. If it does not exist a set of integers $\eta^i$ such that $\omega_i \eta^i = 0$, these trajectories will never close on themselves, and instead will fill the surface of a KAM torus\footnote{See the application of these numerical methods in the context of $\text{AdS}_5 \times \text{T}^{1,1}$ \cite{Basu:2011di} and much more directly related to our studies in \cite{Nunez:2018ags} for the mother $\text{AdS}_7$ backgrounds.}.

The previous results imply that by simply taking cross-sections of the phase-space trajectories we should be able to determine whether or not a system is integrable. As an example, if we plot the pair of data $(\psi_1, G_1)$ each time $\psi_2=\psi_{2,0}$ for $\psi_{2,0}$ an arbitrary number, we will obtain a set of embedded closed KAM trajectories associated to the cross-sections of the embedded KAM tori. This cross-section is the so-called Poincar\'{e} section in the literature. The KAM theorem \cite{Ott:2002book} is a statement about how these KAM trajectories change when we perturb an integrable Hamiltonian system $\mathcal{H}$ by a small deformation $\epsilon \mathcal{H}_{\text{new}}$, with $\epsilon$ denoting a small number. Some of the KAM tori will no longer be closed-shaped curves upon the addition of the deformation. Let assume now, that we increase the strength of the perturbation in the original integrable Hamiltonian system. This will result in the motion becoming completely random and eventually all of the KAM tori will be random shapes. 

To produce Poincar\'{e} sections for the supergravity backgrounds given by \cref{eq: geometry}, we start by first picking a set of appropriate initial conditions. By appropriate we mean that they have to be such that they are different, but have the same energy $E$. We want to make sure that the Virasoro constraint is always satisfied for a given value of the energy, since the Virasoro constraint is a primary condition. We then allow the numerical evolution for the chosen initial points, and plot conjugate pairs of coordinate and momentum at every time.

If the string soliton had an integrable motion, then the phase-space associated trajectories should be confined to some closed-shaped curves, and more specifically tori. The Poincar\'{e} sections would then appear as circles generated by the various tori of different resonances. In the case of the absence of such closed, circular KAM tori, we conclude that the corresponding system is chaotic, and hence non-integrable.

We have already provided concrete evidence based on analytics and numerics that the regular, linear quivers, that can be interpreted as fully localised D$8$-branes in supergravity are chaotic, and thus non-integrable, here we take a complementary approach. 

We will exemplify in a very illustrative way just how special and properly chosen the solution $\alpha(z) = A \sin (\omega z)$ is. To do so, we will present the Poincar\'e sections for this characteristic function. Having done so, we add to that choice of $\alpha(z)$ a very small deformation away from the perfectly chosen sinusoidal shape of the quiver. 

\begin{figure}[ht!]
\centering
\begin{subfigure}[b]{\textwidth}
   \includegraphics[width=1\linewidth]{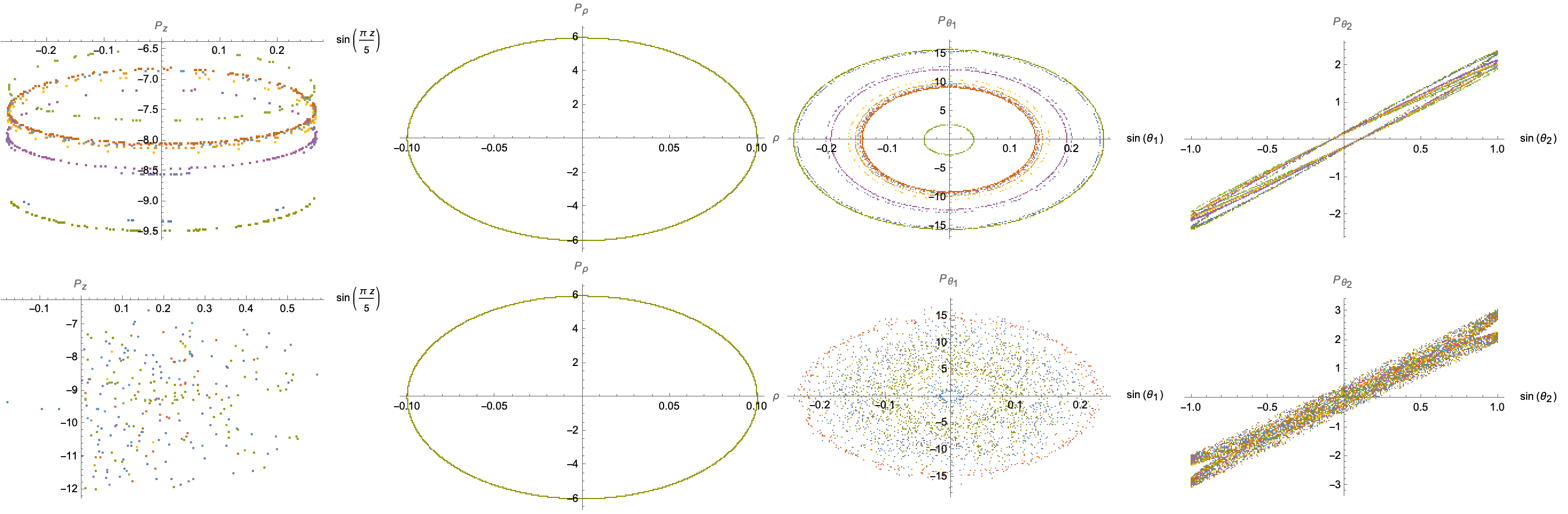}
   \caption{}
   \label{fig: poincare_m23} 
\end{subfigure}

\begin{subfigure}[b]{\textwidth}
   \includegraphics[width=1\linewidth]{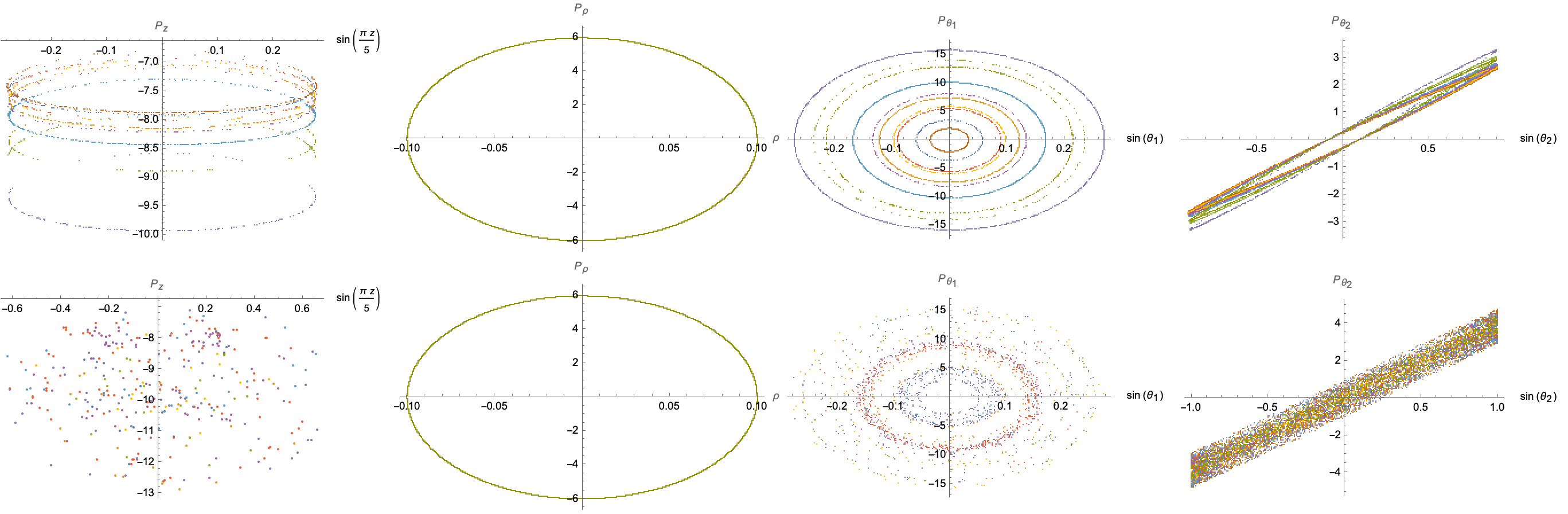}
   \caption{}
   \label{fig: poincare_m34}
\end{subfigure}

\caption{Some indicative Poincar\'{e} sections}
\label{fig: poincare_sections_2334}
{
\Cref{fig: poincare_m23}: Poincar\'{e} sections for the quiver with defining function $\alpha(z) = \sin \left(\tfrac{\pi}{10} z\right)$ on top and for $\alpha(z) = \sin \left(\tfrac{\pi}{10} z\right)+0.00071 z$ on the bottom. The value of the parameter $m$ is $m = \tfrac{2}{3}$. These plots strongly suggest that the classical string motion is integrable in the former case, while a small deformation away from that special solution results in chaotic motion.
\Cref{fig: poincare_m34}: here we have chosen $m=\tfrac{3}{4}$ and $\alpha(z) = \sin \left(\tfrac{\pi}{10} z\right)$ for the top, while $\alpha(z) = \sin \left(\tfrac{\pi}{10} z\right)+0.00071 z$ for the bottom. Regarding the status of (non)-integrability we reach the same conclusions.}
\end{figure}

\begin{figure}[hb!]
\centering
\begin{subfigure}[c]{\textwidth}
   \includegraphics[width=1\linewidth]{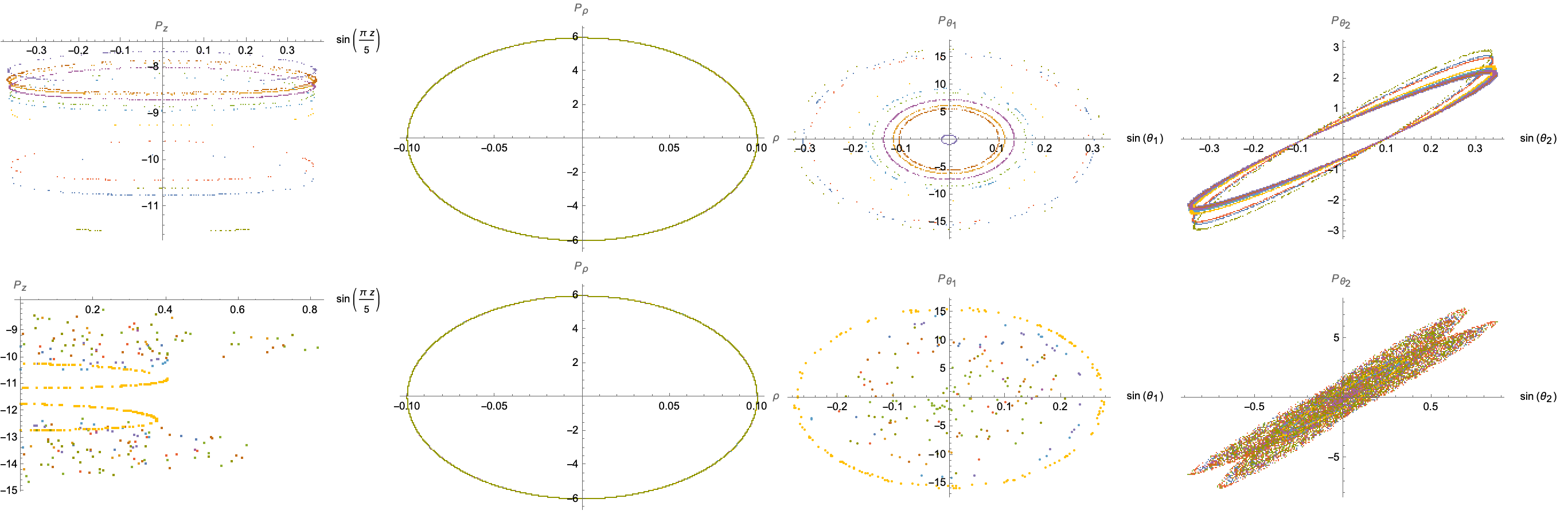}
   \caption{}
   \label{fig: poincare_m1}
\end{subfigure}

\begin{subfigure}[d]{\textwidth}
   \includegraphics[width=1\linewidth]{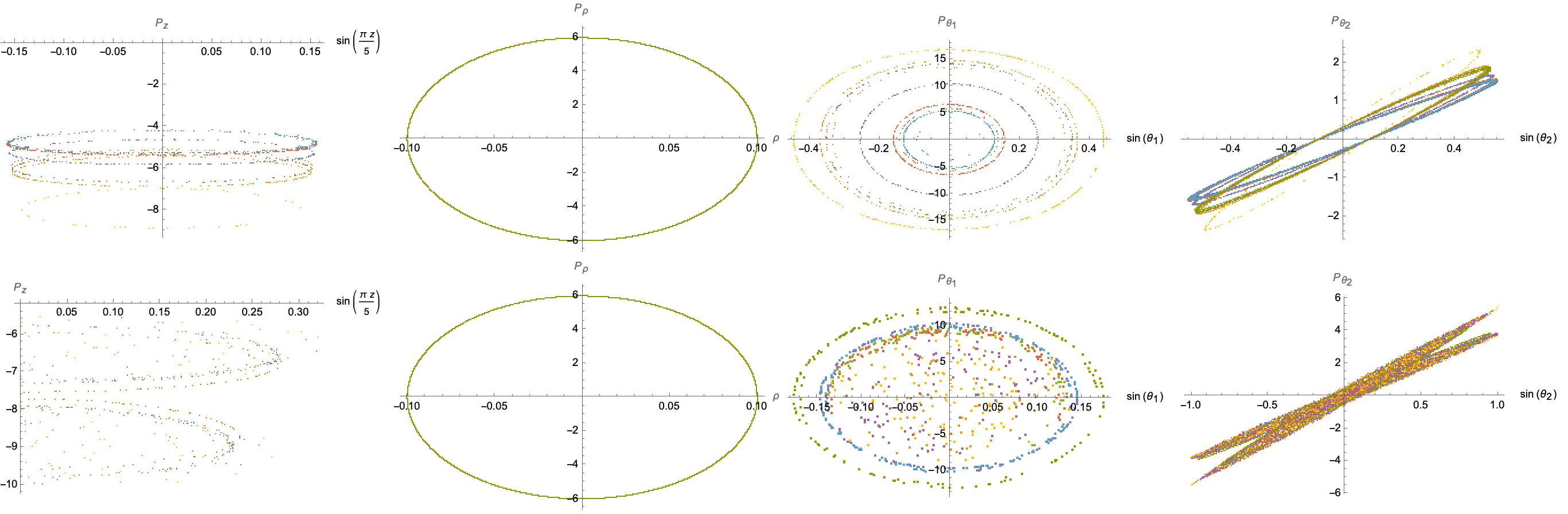}
   \caption{}
   \label{fig: poincare_m179}
\end{subfigure}

\caption{Poincar\'{e} sections}
\label{fig: poincare_sections_1179}
{\Cref{fig: poincare_m1}: Some indicative Poincar\'{e} sections for the quiver with defining function $\alpha(z) = \sin \left(\tfrac{\pi}{10} z\right)$ on top and for $\alpha(z) = \sin \left(\tfrac{\pi}{10} z\right)+0.0953 z$ on the bottom. The value of the parameter $m$ is $m = 1$. These plots strongly suggest that the classical string motion is integrable in the former case, while a small deformation away from that special solution results in chaotic motion. \Cref{fig: poincare_m179}: here we have chosen $m=1.79$ and $\alpha(z) = \sin \left(\tfrac{\pi}{10} z\right)$ for the top, while $\alpha(z) = \sin \left(\tfrac{\pi}{10} z\right)+0.835 z$ for the bottom. Regarding the status of (non)-integrability we reach the same conclusions.} 
\end{figure}

We note that the choice 
\begin{equation}
    \alpha(z) = A \sin (\omega z) + \epsilon \cdot z
\,      ,
\end{equation}
with $\epsilon$ being a small number, does not have a particular physical meaning. It is an interesting case study, though, mathematically. Indeed, as it can be observed from the Poincar\'e sections presented in \cref{fig: poincare_sections_2334,fig: poincare_sections_1179} that even the smallest deformation away from the sinusoidal quiver leads to signatures of chaos. In the Lyapunov exponent spectrum language, this would mean that one exponent is positive and in the power spectrum language it means that the small deformation presents itself as the cause for the trajectories to lose their higher harmonics and the spectrum to be taken over by a band of noise.  

The \cref{fig: poincare_sections_2334,fig: poincare_sections_1179} present some examples that we provide for illustration and completeness. We have checked various other deformations away from the perfect sinusoidal quiver. The same qualitative answer is reached in every single study we performed. 

This leads us to conclude that the choice for the characteristic function $\alpha(z) = A \sin(\omega z)$ is indeed the only choice that does not lead to chaotic string motion. Therefore, it is the only choice for which we can hope the dual field theory to be classically integrable. 

Note, before ending this section, that these numerical findings are in accord with the brief analysis based on the isometries of the background that we presented in \cref{sec: exotic_quiv}. Our argument there was that when $\alpha(z) = A \sin(\omega z)$, the warp factor in front of the $\text{AdS}_5$ submanifold is reduced to a constant and hence our metric becomes the direct product of integrable submanifolds, up to the non-trivial fibration of the geometry. No matter how small a deformation way from that particular solution for $\alpha(z)$ is, the above argument can never hold true, since the warp factors will be non-trivial; even by a small amount. 
\section{Outlook and conclusions}\label{sec: epilogue}
In this work we focused our attention on a recently discovered parametric family of supergravity backgrounds in the context of massive type IIA supergravity that are the dual descriptions of four-dimensional $\mathcal{N}=1$ quiver theories developed in \cite{Merrikin:2022yho}. By performing a careful and detailed study of the dynamics of a string soliton we provide very strong suggestive evidence that the dual field theory is classically integrable only for the choice $\alpha(z) = A \sin(\omega z)$, while all linear quivers are chaotic and hence not Liouville integrable. We have been careful with the choice of the string embedding such that it encapsulates the basic aspects of symmetries in the given class of solutions. Our studies have been both based on analytics and numerics. 

In order to suggest that the sinusoidal quiver leads to an integrable theory, we used Kovacic's algorithm as our main tool with regards to analytics. It is only this choice that can pass all the conditions in all the respective cases. The polynomial rank quivers fail to satisfy said conditions. 

For our numerical approach, we have provided evidence based on no signature of non-integrability and no hallmark behaviour of chaotic system. To do so, we have studied the dynamical evolution of the string in time, and examined how the dynamics are reflected on the plane by the other coordinates. Furthermore, we have computed the spectrum of Lyapunov exponents. Finally, we have studied the power spectra for these differently shaped quivers.

Below, we mention some exciting future directions: 

\begin{itemize}
    \item We have already suggested that the physical interpretation of the special background, defined by $\alpha(z) = A \sin(\omega z)$ is in terms of smeared D$8$-branes. It would be very interesting to find the precise smearing form of this solution using the well-established framework and techniques that have been reviewed in \cite{Nunez:2010sf}.
    \item Our focus here was on the $\mathbb{H}_2$ twisted compactification, however the authors in \cite{Merrikin:2022yho} also studied $\text{T}^2$ and $\text{S}^2$ compactifications. In the latter two cases, however, there were some concerns raised regarding some pathologies of the resulting supergravity solutions. This is the reason that our focus here was on the $\mathbb{H}_2$ since it is well under control. We want to point out, nevertheless, that $\text{T}^2$ and $\text{S}^2$ are maximally symmetric spaces as well and hence an appropriate tuning of the defining function $\alpha(z)$ might lead to integrable string dynamics. For this reason, we believe that it would be very interesting to understand and systematically resolve the issues that were raised in \cite{Merrikin:2022yho}.
    \item It would be very interesting to understand aspects of probe D-branes in both the non-integrable as well as the sinusoidal solution. There is a two-fold motivation for these studies: 
    \begin{enumerate}
        \item Studying supersymmetric probe D-branes in a given background or class of backgrounds is relevant in extracting important information beyond the supergravity regime; namely approaching the stringy regime. This was understood early on in the context of the $\mathcal{N}=4$ SYM in \cite{Witten:1998xy} that the dual gravity picture must contain branes, such that it accommodates the Pfaffian operator for the $\text{SO(N)}$ description or baryon vertices and domain walls in the $\text{SU(N)}$ theory.
        \item Considering the dynamics of any given theory in the presence of boundaries is a very natural way to to probe it. The resulting defect theories are by themselves interesting and, in the past, their studies have led to profound results and insight for the field theories \cite{Gaiotto:2008sd,Gaiotto:2008ak}. 
    \end{enumerate}
    \item It would, also, be very interesting to study and derive the spectrum of spin-2 operators in these backgrounds following the pioneering work of Bachas and Estes \cite{Bachas:2011xa} and see how they fall into different representations of the superconformal algebra. 
    \item In addition to the above, a fascinating direction would be to carefully consider the Penrose limit, derive the corresponding pp-wave geometry \cite{Berenstein:2002jq} of the the massive type IIA solutions and perform the string quantization. 
    \item Another holographic observable of great interest that can be computed is the entanglement entropy \cite{Ryu:2006bv,Kol:2014nqa,Macpherson:2014eza}.  
\end{itemize}

And of course, the three most important, and probably difficult, questions are: 

\begin{itemize}
    \item What is the Lax connection, if one exists, for the sinusoidal solution in these new $\text{AdS}_5$ backgrounds? 
    \item What is so special about these infinite linear quivers that arise in the context of massive type IIA theory that lead to integrability? 
    \item Can we formulate a a no-go theorem for integrability of a string background that has non-trivial warp factors? See \cite{Filippas:2019puw} for related discussion.
\end{itemize}

We hope to be able to report on some of these in the near future. 
\newpage
\section*{Acknowledgments} 
We have benefited from discussions with colleagues and it is a pleasure to thank Carlos Nu\~nez and Riccardo Borsato for interesting suggestions and enlightening comments. KCR is, also, grateful to Kostas Filippas for explanations on Kovacic's algorithm, related to this work, a long time ago, to Huajia Wang for very helpful comments, and to the National Technical University of Athens (NTUA) for the warm hospitality, where this work was finalised. The work of KCR is supported by starting funds from University of Chinese Academy of Sciences (UCAS), the Kavli Institute for Theoretical Sciences (KITS), and the Fundamental Research Funds for the Central Universities.
\newpage
\appendix
\section{The Kovacic algorithm}\label{app: kovacic}
In what follows we summarize, the basic notions of differential Galois theory, which formed the basis used by Kovacic \cite{KOVACIC19863} to construct an algorithm, pertaining to the existence of Liouville integrable solutions on second-order linear ordinary differential equations. By a Liouville integrable, closed-form solution, we are referring to one that is written in terms of algebraic, trigonometric, exponential functions or integrals of those. 

The Kovacic algorithm has been presented in the physics literature in the past; see for example the appendices in \cite{Nunez:2018ags,Filippas:2019ihy,Rigatos:2020igd}. For notational convenience we will follow \cite{Filippas:2019ihy}. We include the Case III of the algorithm, which to the best of our knowledge has not appeared before in the physics literature.

The theorem deals with second-order, linear, ordinary differential equations that assume the form:
\begin{equation}
f''(x)+\mathcal{A}_1(x)f'(x)+\mathcal{A}_2(x)f(x) = 0
\end{equation}
where $x\in\mathbb{C}$ and $\mathcal{A}_{1,2}$ are rational complex functions. We can change variables using $f=e^{\frac{1}{2}\int dx \mathcal{A}_1}z$ to eliminate the first-derivative term and obtain:
\begin{equation}
z''(x)=\mathcal{V}(x)z(x)
\,          ,
\hspace{2cm}
\mathcal{V}=\mathcal{A}_2 -\frac{1}{2} \mathcal{A}^{\prime}_1 - \frac{1}{4} \mathcal{A}^2_1
\,          ,
\label{transformedNVEappendix}
\end{equation}
where $\mathcal{V}$ is called the potential. Obviously, $f$ is Liouvillian integrable if and only if $z$ is, there is no loss of generality under the change of variables.

The starting point of differential Galois theory for these type of equations, the Piccard-Vessiot theory, is the group of automorphisms of its solutions, namely SL$(2,\mathbb{C})$ and its possible subgroups. Let $X$ be an algebraic subgroup of SL$(2,\mathbb{C})$, then one of the four cases can occur:\\

\case{1} \textit{$X$ is triangulisable.}
\\

\case{2} \textit{Case 1 does not hold and}
\\
\textit{$X$ is conjugate to a subgroup of}
\\

\begin{equation}
\left\lbrace\left.\left(\begin{array}{cc}
a &0\\
0 &a^{-1}
\end{array}\right)\right|a\in\mathbb{C}, a\neq0\right\rbrace\cup\left\lbrace\left.\left(\begin{array}{cc}
0 &c\\
-a^{-1} &0
\end{array}\right)\right|a\in\mathbb{C}, a\neq0\right\rbrace
\end{equation}
\hspace{2.1cm}
\\

\case{3} \textit{Case 1 does not hold and the same is true for Case 2 and $X$ is finite.}
\\

\case{4} \textit{$X=$ SL$(2,\mathbb{C})$.}
\\

If the differential equation satisfies one of the three first cases, it has Liouville-integrable solutions. Otherwise, in the case $X=$ SL$(2,\mathbb{C})$, no such solutions exist.
\\

Kovacic managed to translate the first $3$ cases, Cases 1, 2 and 3, into algebraic arguments. These arguments are applied on the behaviour of the potential we defined earlier, $\mathcal{V}$, in \cref{transformedNVEappendix}. These algebraic conditions make up the theorem that is stated as \cite{KOVACIC19863}:\\

\theorem{} \textit{The following necessary conditions have to hold for a particular Case to be true.}\\

\case{1} \textit{Every pole of $\mathcal{V}$ must either have even order or else order 1. The order of $\mathcal{V}$ at $\infty$ must be either even or else greater than 2.}
\\

\case{2} \textit{$\mathcal{V}$ must have at least one pole that either has odd order greater than 2 or else has order exactly 2.}
\\

\case{3} \textit{The order of a pole of $\mathcal{V}$ cannot exceed 2 and the order of $\mathcal{V}$ at $\infty$ must be at least 2.}
\\

If $\mathcal{V}=\texttt{num}/\texttt{denom}$, then the poles of $\mathcal{V}$ are the zeros of $\texttt{denom}$ and the order of the pole is the multiplicity of the zero of $\texttt{denom}$. The order of $\mathcal{V}$ at $\infty$ denotes the number $\deg \texttt{denom}-\deg \texttt{num}$, with $\deg$ denoting the degree of the polynomial.
\\

These conditions are \textit{necessary} for the relevant cases to be true. Ergo, failing these conditions \textit{sufficient} for Case 4 to hold. Hence, we conclude that the failure of all three conditions is enough to declare the differential equation given by \cref{transformedNVEappendix} as Liouville non-integrable.
\\

However, if any of the conditions is satisfied, then the associated Case might hold and, assuming it does, then a Liouville integrable solution exists. So, when a condition is met we have to follow precisely the sub-algorithm of the respective Case to determine if such a solution exists. If a Liouvillian solutions exists, we can use the algorithm to find it. 
\\

Furthermore, Kovacic was able to produce the algorithmic steps for the 3 cases, see \cite{KOVACIC19863}. We present in subsections the algorithm for each case. 
\\

\subsection{The Case 1 algorithm}\label{case1algorithm}
We assume that we are under the condition in which Case 1 holds. We use $\Gamma$ to denote the set of poles of $\mathcal{V}$.\\

\step{1} For each $c\in\Gamma\cup\lbrace\infty\rbrace$ we define two complex
numbers $\alpha_c^\pm$ and a rational function $[\sqrt{\mathcal{V}}]_c$ in the following way:\\

\cc{1} If $c\in\Gamma$ and $c$ is an order 1 pole, then\\
\begin{equation*}
[\sqrt{\mathcal{V}}]_c\:=\:0\hspace{2cm}\alpha_c^\pm\:=\:1
\end{equation*}

\cc{2} If $c\in\Gamma$ and $c$ is an order 2 pole, then\\
\begin{equation*}
[\sqrt{\mathcal{V}}]_c\:=\:0
\end{equation*}\\
Let $\beta_c$ be the coefficient of the term $1/(x-c)^2$ in the expansion in partial fractions for $\mathcal{V}$. Then\\
\begin{equation*}
\alpha_c^\pm\:=\:\frac{1}{2}\pm\frac{1}{2}\sqrt{1+4\beta_c}
\end{equation*}

\cc{3} If $c\in\Gamma$ and $c$ is an order $2\nu\geq4$ pole, which in Case 1 is necessarily an even number, then $[\sqrt{\mathcal{V}}]_c$ is the sum of all the terms that involve $1/(x-c)^i$ for $2\leq i\leq\nu$ in the Laurent series expansion of $\sqrt{\mathcal{V}}$ at $c$. There are two possibilities for $[\sqrt{\mathcal{V}}]_c$, that are equal up to the overall sign, and we can choose either one of them. Hence\\
\begin{equation*}
[\sqrt{\mathcal{V}}]_c\:=\:\frac{a}{(x-c)^\nu}+\dots+\frac{d}{(x-c)^2}
\end{equation*}\\
Let $\beta_c$ be the coefficient of the term $1/(x-c)^{\nu+1}$ in $\mathcal{V}$ minus the coefficient of the term $1/(x-c)^{\nu+1}$ in $[\sqrt{\mathcal{V}}]^2_c$. We have\\
\begin{equation*}
\alpha_c^\pm\:=\:\frac{1}{2}\left(\pm\frac{\beta_c}{a}+\nu\right)
\end{equation*}\\

\cci{1} If the order of $\mathcal{V}$ at $\infty$ is bigger than $2$, then\\
\begin{equation*}
[\sqrt{\mathcal{V}}]_\infty\:=\:0\hspace{2cm}\alpha_\infty^+\:=\:0\hspace{1cm}\alpha_\infty^-\:=\:1
\end{equation*}

\cci{2} If the order of $\mathcal{V}$ at $\infty$ is equal to 2, then\\
\begin{equation*}
[\sqrt{\mathcal{V}}]_\infty\:=\:0
\end{equation*}\\
Let $b_\infty$ be the coefficient of the term $1/x^2$ in the Laurent series expansion of $\mathcal{V}$ at $\infty$. (If $\mathcal{V}=\texttt{num}/\texttt{denom}$, where $\texttt{num}$, $\texttt{denom}$ are coprime, then $b_\infty$ is the leading coefficient of $\texttt{num}$ divided by the leading coefficient of $\texttt{denom}$.). Then\\
\begin{equation*}
\alpha_\infty^\pm\:=\:\frac{1}{2}\pm\frac{1}{2}\sqrt{1+4\beta_\infty}
\end{equation*}

\cci{3} If the order of $\mathcal{V}$ at $\infty$ is $-2\nu\leq0$, which in Case 1 is necessarily an even number, then $[\sqrt{\mathcal{V}}]_\infty$ is the sum of all terms involving $x^i$ for $0\leq i\leq\nu$ in the Laurent series for $\sqrt{\mathcal{V}}$ at $\infty$. We can choose either of the two possibilities. Ergo\\
\begin{equation*}
[\sqrt{\mathcal{V}}]_\infty\:=\:ax^\nu+\dots+d
\end{equation*}\\
Let $\beta_\infty$ be the coefficient of $x^{\nu-1}$ in $\mathcal{V}$ minus the coefficient of $x^{\nu-1}$ in $([\sqrt{\mathcal{V}}]_\infty)^2$. Then\\
\begin{equation*}
\alpha_\infty^\pm\:=\:\frac{1}{2}\left(\pm\frac{\beta_\infty}{a}-\nu\right)
\end{equation*}\\

\step{2} For each family $\texttt{num}=(\texttt{num}(c))_{c\in\Gamma\cup\lbrace\infty\rbrace}$, where $\texttt{num}(c)$ is $+$ or $-$, let\\
\begin{equation*}
d\:=\:\alpha_\infty^{s(\infty)}-\sum_{c\in\Gamma}\alpha_c^{s(c)}
\end{equation*}\\
If $d$ is a non-negative integer, then\\
\begin{equation*}
\omega\:=\:\sum_{c\in\Gamma}\left(s(c)[\sqrt{\mathcal{V}}]_c+\frac{\alpha_c^{s(c)}}{x-c}\right)+s(\infty)[\sqrt{\mathcal{V}}]_\infty
\end{equation*}\\
is a candidate for $\omega$. If $d$ is not a non-negative integer, then the family $\texttt{num}$ can be removed from consideration.\\

\step{3} We have to apply this step to each one of the families retained from Step 2, until we succeed or all of the families have been exhausted. In the latter case, Case 1 does not hold.\\[5pt]
For each family, we search for a monic polynomial $P$ of degree $d$ (as defined in Step 2) that satisfies the differential equation\\
\begin{equation*}
P''+2\omega P'+(\omega'+\omega^2-\mathcal{V})P\:=\:0
\end{equation*}\\
This we can achieve conveniently by using the undetermined coefficients and is a simple linear-algebra problem. Note that the problem may or may not have a solution. If such a polynomial exists, then $\eta=Pe^{\int\omega}$ is a solution of the differential equation \cref{transformedNVEappendix}. If there is no such polynomial for any family retained from Step 2, then Case 1 does not hold.
\\

\subsection{The Case 2 algorithm}\label{case2algorithm}
In order for this algorithm to take effect, we assume that Case 1 fails. We work similarly to Case 1, namely we first gather the data for each pole $c$ of $\mathcal{V}$ and also for $\infty$. The data form a set that is either $E_c$ or $E_\infty$ and consists of some integer numbers, from one to three. Next, we consider the families of elements of these sets, discarding some and retaining others. If no families are retained, Case 2 does not hold. However, for each family that is retained we search for a monic polynomial that obeys a certain linear differential equation. If no such polynomial exists for any family, then Case 2 does not hold. If such a polynomial exists, then we have obtained a solution to the differential equation given by \cref{transformedNVEappendix}.\\

We use $\Gamma$ to denote the set of poles of $\mathcal{V}$.\\

\step{1} For each $c\in\Gamma$ we define $E_c$ as follows.\\

\cc{1} If $c$ is an order 1 pole, then $E_c=\lbrace4\rbrace$.\\

\cc{2} If $c$ is an order 2 pole and if $\beta_c$ is the coefficient of $1/(x-c)^2$ in the expansion in partial fractions of $\mathcal{V}$, then
\begin{equation*}
E_c=\lbrace 2+k\sqrt{1+4\beta_c}|k=0,\pm2\rbrace\cap\mathbb{Z}
\end{equation*}

\cc{3} If $c$ is an order $\nu>2$ pole, then $E_c=\lbrace\nu\rbrace$.\\

\cci{1} If $\mathcal{V}$ has order bigger than $2$ at $\infty$ , then $E_\infty=\lbrace0,2,4\rbrace$.\\

\cci{2} If $\mathcal{V}$ has order exactly 2 at $\infty$ and $\beta_\infty$ is the coefficient of $\mathcal{V}$ in the Laurent series expansion of $\mathcal{V}$ at $\infty$, then
\begin{equation*}
E_\infty=\lbrace 2+k\sqrt{1+4\beta_\infty}|k=0,\pm2\rbrace\cap\mathbb{Z}
\end{equation*}

\cci{3} If the order of $\mathcal{V}$ at $\infty$ is $\nu<2$, then $E_\infty=\lbrace\nu\rbrace$.\\

\step{2} We account for all families $(e_c)_{c\in\Gamma\cup\lbrace\infty\rbrace}$ with $e_c\in E_c$. Any family that has all its coordinates to be even can be discarded. Consider\\
\begin{equation*}
d\:=\:\frac{1}{2}\left(e_\infty-\sum_{c\in\Gamma}e_c\right)
\end{equation*}\\
If $d$ is a non-negative integer, then the family should be retained. If not, then the family is discarded. If there are no families retained, then Case 2 does not hold.\\

\step{3} For each of the families retained from Step 2, we construct the rational function\\
\begin{equation*}
\theta\:=\:\frac{1}{2}\sum_{c\in\Gamma}\frac{e_c}{x-c}
\end{equation*}\\
We next search for a monic polynomial $P$ of degree $d$, as defined in Step 2, such that it satisfies\\
\begin{equation*}
P'''+3\theta P''+(3\theta^2+3\theta'-4\mathcal{V})P'+(\theta''+3\theta\theta'+\theta^3-4\mathcal{V}\theta-2\mathcal{V}')P\:=\:0
\end{equation*}\\
If we cannot find such a polynomial for any of the families retained from Step 2, then case 2 does not hold.\\[5pt]
Let us assume that we have found such a polynomial. Let us define $\varphi=\theta+P'/P$ and let $\omega$ be a solution of the equation
\begin{equation*}
\omega^2+\varphi\omega+(\frac{1}{2}\varphi'+\frac{1}{2}\varphi^2-\mathcal{V})\:=\:0
\end{equation*}\\
Then $\eta=e^{\int\omega}$ is a solution of the differential equation given by \cref{transformedNVEappendix}.\\

\subsection{The Case 3 algorithm}\label{case3algorithm}
In Case 3, we assume that Cases 1 and 2 fail and the differential equation only has algebraic solutions. Let us assume $\eta$ is a solution of the differential equation $z^{\prime \prime} = \mathcal{V} z$ and let us define $\omega = \tfrac{\eta^{\prime}}{\eta}$. In this case, $\omega$ is algebraic over $\mathbb{C}(x)$ of degree 4, 6, or 12 and it is the minimal polynomial for $\omega$ that we will have to find. 

To do so, there are two possible ways. On the one hand, we can find a polynomial of degree 12 and then factor it. If $\omega$ is any solution of he 12th degree polynomial, then $\eta = e^{\int \omega}$ is a solution to the differential equation, and hence any of the irreducible factors can be used. On the other hand, we can first attempt to find a 4th degree equation for $\omega$, subsequently a 6th degree equation and finally a 12th degree equation. If an equation is found, it is guaranteed to be irreducible. 

Let $\Gamma$ be the set of poles of $\mathcal{V}$. Recall that by the necessary conditions of Case 2, $\mathcal{V}$ cannot have a pole of order higher than 2. \\

\step{1} For each $c\in\Gamma\cup\lbrace\infty\rbrace$ we define $E_c$ as follows.\\

\cc{1} If $c$ is an order 1 pole, then $E_c=\lbrace12\rbrace$.\\

\cc{2} If $c$ is an order 2 pole and if $\alpha_c$ is the coefficient of the term $1/(x-c)^2$ in the expansion of $\mathcal{V}$ in partial fractions, then
\begin{equation*}
E_c=\lbrace 6+\frac{12k}{n}\sqrt{1+4\alpha_c}|k=0,\pm1,\pm2,\ldots,\pm\frac{n}{2}\rbrace\cap\mathbb{Z}
\end{equation*}

\cci{} If the Laurent series for $\mathcal{V}$ at $\infty$ has the form\\
\begin{equation*}
\mathcal{V} = \gamma x^{-2}+\ldots\,        ,       \qquad      (\gamma \in \mathbb{C}, \quad \text{possibly} \quad0)        
\,      ,
\end{equation*}\\
then\\
\begin{equation*}
E_\infty=\lbrace 6+\frac{12k}{n}\sqrt{1+4\gamma}|k=0,\pm1,\pm2,\ldots,\pm\frac{n}{2}\rbrace\cap\mathbb{Z}
\end{equation*}\\

\step{2} We consider all families $(e_c)_{c\in\Gamma\cup\lbrace\infty\rbrace}$ with $e_c\in E_c$. For each family define
\\
\begin{equation*}
d = \frac{n}{12}\left(e_\infty-\sum_{c\in\Gamma}e_c\right)
\,          .
\end{equation*}
\\
If $d$ is a non-negative integer, the should retain the family, in any other case the family has to be discarded. If no families remain under consideration, then Case 3 does not hold.
\\

\step{3} For each of the families from Step 2, we construct the rational function
\\
\begin{equation*}
\theta=\frac{n}{2}\sum_{c\in\Gamma}\frac{e_c}{x-c}
\end{equation*}
\\
And we, further, construct
\\
\begin{equation*}
S = \prod_{c\in\Gamma}(x-c)
\end{equation*}
\\

Next we look for a monic polynomial $P \in \mathbb{C}(x)$ that has degree $d$ (as defined in step 2) in such a way that when we compute the polynomials $P_n, P_{n-1},\ldots,P_{-1}$ in a recursive manner by the fomulae:
\\
\begin{equation*}
\begin{aligned}
P_n &= - P
\,          ,
\\
P_{i-1} &= -SP^{\prime}_i + ((n-i)S^{\prime} - S \theta)P_i - (n-i)(i+1)S^2 r P_{i+1}
\,          ,
\\
\text{with} \qquad &i=n,n-1,\ldots,0
\,          ,
\end{aligned}   
\end{equation*}
\\
then $P_{-1}=0$ identically. 
\\

If we cannot find such a polynomial for any of the families retained from step 2, then case 3 does not hold.
\\

Let us assume that we have found such a polynomial. Consider $\varphi=\theta+P'/P$ and let $\omega$ be a solution of the equation:
\begin{equation*}
\sum^{n}_{i=0} \frac{S^i P_i}{(n-i)!} \omega^i = 0
\,          .
\end{equation*}
\\

Then $\eta=e^{\int\omega}$ is a solution of \cref{transformedNVEappendix}.
\\
\newpage
\bibliographystyle{ssg}
\bibliography{integrablequiver}
\end{document}